\date{\today}

\documentclass[aps,prb,twocolumn,superscriptaddress,groupedaddress]{revtex4}  % for review and submission
\usepackage{dcolumn,graphicx,amsmath,amssymb,algorithm,algpseudocode}
\usepackage{todonotes}
\usepackage{qcircuit}

\newcommand{\total}{\mathrm{d}}
\newcommand{\ud}{\mathrm{d}}

\newcommand{\diff}[2]{\frac{\ud {#1}}{\ud {#2}}}
\newcommand{\pdiff}[2]{\frac{\partial #1}{\partial #2}}

\begin{document}

\definecolor{brickred}{rgb}{.72,0,0} 

\title{
Hybrid Quantum/Classical Derivative Theory:
Analytical Gradients and Excited-State Dynamics for the Multistate Contracted
Variational Quantum Eigensolver
}

\author{Robert M. Parrish}
\email{rob.parrish@qcware.com}
\affiliation{
QC Ware Corporation, Palo Alto, CA 94301
}
\affiliation{
Department of Chemistry and the PULSE Institute, Stanford University, Stanford, CA 94305
    }
\affiliation{
SLAC National Accelerator Laboratory, Menlo Park, CA 94025
}

\author{Edward G. Hohenstein}
\affiliation{
Department of Chemistry and the PULSE Institute, Stanford University, Stanford, CA 94305
    }
\affiliation{
SLAC National Accelerator Laboratory, Menlo Park, CA 94025
}

\author{Peter L. McMahon}
\affiliation{
QC Ware Corporation, Palo Alto, CA 94301
}
\affiliation{
E.\,L.~Ginzton Laboratory, Stanford University, Stanford, CA 94305
    }

\author{Todd J. Mart\'inez}
\affiliation{
Department of Chemistry and the PULSE Institute, Stanford University, Stanford, CA 94305
    }
\affiliation{
SLAC National Accelerator Laboratory, Menlo Park, CA 94025
}

\begin{abstract} 
The maturation of analytical derivative theory over the past few decades has
enabled classical electronic structure theory to provide accurate and
efficient predictions of a wide variety of observable properties. However,
classical implementations of analytical derivative theory take advantage of
explicit computational access to the approximate electronic wavefunctions in
question, which is not possible for the emerging case of hybrid
quantum/classical methods. Here, we develop an efficient Lagrangian-based
approach for analytical first derivatives of hybrid quantum/classical methods
using only observable quantities from the quantum portion of the algorithm.
Specifically, we construct the key first-derivative property of the nuclear
energy gradient for the recently-developed multistate, contracted variant of the
variational quantum eigensolver (MC-VQE) within the context of the \emph{ab
initio} exciton model (AIEM). We show that a clean separation between the
quantum and classical parts of the problem is enabled by the definition of an
appropriate set of relaxed density matrices, and show how the wavefunction
response equations in the quantum part of the algorithm (coupled-perturbed
MC-VQE or CP-MC-VQE equations) are decoupled from the wavefunction response
equations and and gradient perturbations in the classical part of the algorithm.
We explore the magnitudes of the Hellmann-Feynman and response contributions to
the gradients in quantum circuit simulations of MC-VQE+AIEM and demonstrate a
quantum circuit simulator implementation of adiabatic excited state dynamics
with MC-VQE+AIEM.
\end{abstract}

% 03.67.Ac	Quantum algorithms, protocols, and simulations
% 31.10.+z	Theory of electronic structure, electronic transitions, and chemical binding
% 31.15.-p	Calculations and mathematical techniques in atomic and molecular physics 
% 03.65.-w	Quantum mechanics
% \pacs{03.67.Ac,31.10.+z,31.15.-p}

\maketitle

\section{Introduction}

The emergence of hybrid variational quantum/classical algorithms\cite{
Peruzzo:2014:4213}
for the
approximate diagonalization of the electronic Hamiltonian represents a promising
pathway to the robust and accurate determination of observable properties in
strongly-correlated molecular systems on noisy intermediate-scale quantum (NISQ)
hardware.\cite{Preskill:2018:79} However, most efforts to this point have
focused on ground-state\cite{
Peruzzo:2014:4213,
McClean:2016:023023,
OMalley:2016:031007,
Kandala:2017:242,
McClean:2017:X,
Romero:2018:104008,
Nam:2019:IonWater}
and, more-recently, excited-state energies\cite{
Peruzzo:2014:4213,
McClean:2017:042308,
Higgott:2018:X,
Lee:2018:JCTC,
Colless:2018:011021,
Nakanishi:2018:VQE,
Parrish:2019:230401}
at a single nuclear geometry. To make further progress, efforts are needed to
extend these methods to the efficient computation of analytical derivative
properties such as nuclear energy
gradients,\cite{
Bratoz:1958:RHFgrad,
Gerratt:1968:Force,
Pulay:1969:SCFgrad,
Kato:1979:MCSCFgrad,
Goddard:1979:TCSCFgrad,
Pulay:1979:Systematic,
Brooks:1980:Analytic,
Krishnan:1980:CIDderivative,
Dupuis:1981:Energy,
Nakatsuji:1981:Force,
Schaefer:1986:new}
relaxed dipole moments,\cite{
Diercksen:1981:29, 
Almlof:1985:743} 
dipole derivatives,\cite{
Gerratt:1968:Force, 
Pulay:1979:Systematic, 
Yamaguchi:1986:2262} 
electronic polarizabilities,\cite{
Ball:1964:501, 
Mclachlan:1964:844, 
Cohen:1965:S34, 
Hurst:1988:385} 
polarizability derivatives,\cite{
Amos:1986:376} 
circular dichroism spectra,\cite{
Stephens:1994:CD,Mccann:2006:CD,Crawford:2007:CD}
nuclear magnetic resonance (NMR) shielding tensors,\cite{
Wolinski:1990:8251,
Keith:1992:1,
Keith:1993:223,
Ruud:1993:3847,
Cheeseman:1996:5497,
Helgaker:1999:ab,
Ochsenfeld:2004:NMR} 
NMR spin-spin coupling constants,\cite{
Sychrovsky:2000:3530,
Helgaker:2000:9402}
hyperfine coupling constants,\cite{
Fernandez:1992:Spin} 
vibrational frequencies,\cite{
Pulay:1969:SCFgrad,
Pople:1981:Hessian,
Osamura:1982:Unified,
Pulay:1979:Systematic} 
and non-adiabatic coupling vectors.\cite{
Lengsfield:1984:4549,
Saxe:1985:159,
Lengsfield:1986:348,
Saxe:1987:321,
Lengsfield:1992:1}  
In attempting such extensions, we will have to contend with the fact that the
hybrid quantum/classical methods generally utilize an approximate wavefunction
ansatz and will therefore carry wavefunction response terms in the desired
analytical derivative properties. The efficient computation of such analytical
derivative properties using only the simple low-order Pauli expectation values
that are available at the output of quantum circuits (i.e., without explicit
knowledge of the many-electron wavefunctions) is the major focus of the present
work.  Specifically, we will focus on developing analytical nuclear gradients of
our recently-developed multistate, contracted variant of the variational quantum
eigensolver\cite{Parrish:2019:230401} (MC-VQE) within the ab initio exciton
model\cite{
Sisto:2014:2857,
Sisto:2017:14924,
Li:2017:3493}
(AIEM) framework. MC-VQE provides a route to the balanced treatment of
ground-state, excited-state and transition properties, and is the latest in an
extensive series of methods proposed to extend VQE to accurately and efficiently
handle excited states.\cite{
Peruzzo:2014:4213,
McClean:2017:042308,
Higgott:2018:X,
Lee:2018:JCTC,
Colless:2018:011021,
Nakanishi:2018:VQE}

In the context of classical electronic structure theory, much progress has been
made in the last few decades in the computation of accurate observable
properties as analytical derivatives of expectation values of approximate
electronic wavefunctions. As one key example, determining the analytical
gradient of the adiabatic energy for a given electronic state with respect to
the nuclear positions yields the classical forces acting on the nuclei at a
given nuclear configuration and electronic state. This provides first-order
Taylor series information of the potential energy surface for the electronic
state, and allows for myriad applications that would be otherwise intractable
with only energies.\cite{jorgensen2012geometrical} For instance, gradient-based
optimization of the energy with respect to nuclear coordinates yields local
energy minima of the potential energy surface, which are useful proxies for the
stable equilibrium geometries of molecular species.  First-order saddle points
on the potential energy surface can be located with similar gradient-based
algorithms, yielding the transition state structures for interesting chemical
reactions. Second-order derivatives of the energy with respect to the nuclear
coordinates (Hessians) can be used at optimal structures to obtain a harmonic
approximation for the nuclear wavepacket, including predictions of zero-point
vibrational energy, finite-temperature vibrational enthalpy/entropy, and
infrared and Raman spectra.  Taken together, the first and second-order
derivative properties at two minimal structures and the connecting transition
state can provide an estimate of the reaction energy and a
transition-state-theory estimate of the reaction rate.\cite{mills1994quantum,henkelman2000climbing,weinan2002string,peters2004growing,behn2011efficient}  Additionally, in the
real time axis, analytical nuclear gradients can also be used to perform
\emph{ab initio} molecular dynamics (AIMD), which, in its simplest form,
involves the classical Newtonian propagation of the nuclei along the
Born-Oppenheimer electronic potential surface. Note that with efficient codes
for the computation of the electronic energy and its analytical nuclear gradient
at a given nuclear configuration, AIMD can be performed with ``on-the-fly''
sampling of the potential energy surface, allowing for its deployment in systems
with thousands of atoms, e.g., no intractable grid-based representation of the
potential surface is needed. AIMD is particularly useful for computing
time-resolved observables of non-equilibrium chemical processes.  Many
extensions of AIMD have been developed to account for the fact that the nuclei
are non-classical and/or to relax the Born-Oppenheimer approximation and target
the full solution of the molecular Schr\"odinger equation - such methods include
ab initio multiple spawning (AIMS),\cite{ben-nun_ab_2000} fewest switches
surface hopping (FSSH),\cite{tully_molecular_1990} multi-configurational
Ehrenfest (AI-MCE),\cite{saita_--fly_2012} variational multiconfiurational
Gaussian methods (vMCG),\cite{richings_quantum_2015} and many variants. Aside
from heavy reliance on the nuclear energy gradient, a critical derivative
quantity encountered ubiquitously in AIMS and the other non-adiabatic dynamics
methods is the ``non-adiabatic coupling vector,'' which determines how much the
electronic overlap between two adiabatic states changes as one of the states is
moved.\cite{
Lengsfield:1984:4549,
Saxe:1985:159,
Lengsfield:1986:348,
Saxe:1987:321,
Lengsfield:1992:1}  
The non-adiabatic coupling vector is usually formulated as a highly
unusual derivative property that does not straightforwardly resemble an
observable but nonetheless relies heavily on analytical derivative machinery -
we will discuss the hybrid quantum/classical treatment of the non-adiabatic
coupling vector in MC-VQE in a forthcoming companion paper. 

In all of these applications, the practitioners of analytical gradient theory
have developed two coupled governing principles: 
\begin{enumerate}
\item Regardless of the origins or definitions of the approximate electronic
wavefunction, one should always take the \emph{exact} derivative of the
approximated observable expectation value - no further approximations should be
permitted in the derivative. This makes the derivative self-consistent with the
approximated observable expectation value, which often provides for markedly
favorable cancellation of errors in properties. In many cases, this principle is
fundamentally required for practical use the derivative property: As one such
instance, the gradient of the energy with respect to nuclear coordinates must be
self-consistent with the approximate energy observable to allow for the equations of
motion of \emph{ab initio} molecular dynamics to be integrated while respecting
the inviolable invariant of conservation of energy. Generally, satisfying this
principle mandates the consideration of the derivatives of the approximations
built into the wavefunction definitions with respect to gradient perturbations,
a topic loosely known as ``wavefunction response.''
\item With careful effort, it is generally possible to formulate first
derivatives of arbitrary observable expectation values in a way that depends
weakly (or ideally not at all) on the number of gradient perturbations. More
concretely, it is generally possible to restructure the derivative problem in a
way where the explicit wavefunction response does not need to be computed
separately for each gradient perturbation - instead, an effective collective
wavefunction response can be computed once and used in conjunction with the
chain rule to efficiently compute the desired total derivatives. Often this
leads to the ideal case where the computation of the analytical derivative
property costs the \emph{same} as the underlying observable, to within some
constant prefactor.
\end{enumerate}
Analytical derivative theory has a long and rich history in the electronic
structure literature. Starting from the pioneering work by Pulay and others in the 
computation of the analytical gradient of approximate Hartree-Fock theory,\cite{
Bratoz:1958:RHFgrad,
Gerratt:1968:Force,
Pulay:1969:SCFgrad} it
was immediately noticed that analytical derivative theory was plagued by extreme
verbosity of the required equations, requiring careful efforts to produce
correct derivatives (particularly those involving explicit wavefunction response
terms). The explicit forward differentiation of the wavefunction response
contributions was heavily developed during the 1980s in a style characterized by
noteworthy contributions from Yamaguchi and Schaefer.\cite{schaefer1986new} At
around this time, a major breakthrough was realized in the widespread deployment
of Handy-Schaefer $Z$-vector method\cite{Handy:1984:Zvec} (also sometimes known
as the ``Delgarno-Stewart interchange theorem''\cite{Dalgarno:1958:245}), which
removed the need to explicitly solve for the response of the wavefunction
parameters to each gradient perturbation.  The $Z$-vector method substantially
accelerated analytical gradient theory, to the point that it was generally far
superior to finite difference approximations in both runtime and accuracy, but
was often seen as a clever mathematical/computational manipulation rather than a
fundamental feature of analytical gradient theory. This changed with the
widespread adoption of the Lagrangian formalism of
Helgaker\cite{Helgaker:1982:LAG} for analytical derivatives of approximate
wavefunctions in the late 1980s and early 1990s\cite{
Jorgensen:1988:MP2Lag,
Helgaker:1989:Numerically,
Helgaker:1989:Configuration,
helgaker1992calculation,
Fernandez:1992:Spin,
Szalay:1995:Analytic,
Helgaker:1999:ab,
Staalring:2001:Analytical,
Hald:2003:Lagrangian,
Levchenko:2005:Analytic,
Coriani:2010:Atomic,
Rekkedal:2013:Communication} - the non-variational energy (or other
observable) expression for a given approximate wavefunction method can be
exchanged for an equivalent Lagrangian scalar quantity with additional Lagrange
multiplier parameters. Making the Lagrangian variational with respect to the
Lagrange multiplier parameters provides a succinct and rigorous definition of
the usual wavefunction parameters, while making the Lagrangian variational with
respect to the usual wavefunction parameters determines the values of the
Langrange multiplier parameters through a series of linear ``wavefunction
response'' equations.  Notably, the Handy-Schaefer $Z$-vector method arises
naturally in the Lagrangian formalism, automatically minimizing the number of
wavefunction response equations that must be solved for a given observable
(regardless of the number of derivative perturbations).

While there have been considerable recent efforts to develop hybrid
quantum/classical methods for zeroth-order scalar observables such as
ground-state and excited-state energies and transition properties, the existing
literature on hybrid quantum/classical analytical derivative theory is notably
sparse. An approach for the nuclear energy gradient (and higher derivatives) has
been proposed\cite{Kassal:2009:grad} based on Jordan's quantum gradient
estimation algorithm within the phase estimation algorithm. More recently,
another approach has been proposed\cite{Roggero:2018:linear} for linear
response based on perturbations of the phase estimation algorithm.  Within the
original VQE method, it is also clear that the nuclear energy gradient of the
ground state is straightforward to compute, as the variational density matrix
is available as a byproduct of the VQE optimization.\cite{McClean:2016:023023}
However, this approach will not be applicable to most excited state VQE
extensions or to transition properties, as here the quantities to be
differentiated are not variational in the quantum circuit parameters. Finally,
as the numerical experiments in this  manuscript were being finalized, two
separate groups have proposed methodology for the computation of the nuclear
energy gradient using a sum-over-states approach\cite{Obrien:2019:grad} and
using an explicit differentiation approach.\cite{Mitarai:2019:grad} Both groups
demonstrate their method in the context of H$_2$. Notably, no correspondence of
the Handy-Schaefer $Z$-vector method or the Lagrangian formalism has yet been
introduced - existing approaches either ignore the effect of wavefunction
response or compute it through direct forward evaluation of the response
derivatives\cite{Mitarai:2019:grad} or effectively through a sum-over-states
resolution.\cite{Obrien:2019:grad}

In the present manuscript, we first review some technical prerequisites related
to the Lagrangian formalism of analytical derivative theory and
efficient/accurate techniques for computing the analytical gradients of quantum
circuit observable expectation values with respect to circuit parameters. We then
carefully define and differentiate each stage of the MC-VQE algorithm in the
specific context of the AIEM. The expressions developed herein are deliberately
specific to the AIEM to provide an impression of the flow of a Lagrangian
workflow for hybrid quantum/classical derivative theory. However, the overall
steps would be similar for other Hamiltonian representations, such as
fermionic systems represented by the
Jordan-Wigner,\cite{Jordan:1928:631,Ortiz:2001:022319}
Bravyi-Kitaev,\cite{Bravyi:2002:210,Seeley:2012:224109} or other spin-lattice
representations.\cite{Setia:2017:X,Setia:2018:X,Babbush:2017:X,
Kivlichan:2018:110501,Motta:2018:X} Moreover, the general hybrid
quantum/classical Lagrangian approach adopted herein should be straightforward to
adapt to other variants of VQE. Finally, the broad sketches of the approach
developed in this work should apply equally to other derivative properties such
as non-adiabatic coupling vectors - in particular, the various response
equations appearing here will be identical up to the choice of right-hand side.

\section{Technical Background}

\subsection{Analytical Derivative Theory} 

Analytical derivative theory is concerned with the computation of derivatives of
expectation values of observable quantities of approximate wavefunctions,
\begin{equation}
\total_{\zeta}
O_{\Theta}
\equiv
\total_{\zeta}
\langle \Psi_{\Theta} | \hat O | \Psi_{\Theta} \rangle
\end{equation}
Here $\total_{\zeta}$ is shorthand for the total derivative $\total / \total
\zeta$ ($\partial_{\zeta}$ will similarly serve as shorthand for the partial
derivative $\partial / \partial \zeta$). For instance, substituting $\hat O
\equiv \hat H$ and taking $\zeta$ to be the Cartesian coordinates of the nuclei
yields the traditional ``energy gradient,'' $E^{\zeta}$ which is equivalent to
the opposite of the force acting on the nuclei at a given nuclear geometry. We
will specialize to the case of the nuclear gradient in all derivations in the
present manuscript. Analytical derivatives of other properties such as
non-adiabatic coupling vectors, dipole derivatives, etc, would follow similar
manipulations.

In general there are several contributions to the gradient,
\begin{equation}
\total_{\zeta}
E_{\Theta}
=
\langle \Psi_{\Theta} | \total_{\zeta} \hat H | \Psi_{\Theta} \rangle
+
\langle \Psi_{\Theta} | \hat H | \total_{\zeta} \Psi_{\Theta} \rangle
+
\mathrm{H.C.}
\end{equation}
The first term is the ``Hellmann-Feynman''
contribution,\cite{guttinger1932verhalten,pauli1933principles,hellman1937einfuhrung,feynman1939forces}
which reflects the expectation value of the intrinsic derivative of the
Hamiltonian with respect to $\zeta$. The last two terms are the ``wavefunction
response'' contributions, which reflect the fact that the wavefunction
parameters may vary with respect to $\zeta$, providing an additional nonzero
contribution to the gradient. 
    
More explicitly, the wavefunction $|\Psi_{\Theta} \rangle$ might definitionally
depend on a set of parameters $\{ \theta_{i} \}$, which themselves depend on
$\hat H$ (or perhaps other operators), and from thence depend on $\zeta$. E.g.,
$\{ \theta_i \}$ might be determined by solving another electronic structure
method involving implicit equations in $\hat H$. Therefore,
\begin{equation}
\diff{E_{\Theta} (\hat H, \{ \theta_{i} \} )
}{
\zeta}
=
\pdiff{E_{\Theta}}{\hat H}
\diff{\hat H}{\zeta}
+
\pdiff{E_{\Theta}}{\theta_i}
\diff{\theta_i}{\zeta}
\end{equation}
The explicit computation of the wavefunction response over a large number of
perturbation coordinates $\zeta$ may prove to be exhaustingly tedious.

The Lagrangian formalism\cite{Helgaker:1982:LAG} can help overcome this
difficulty and reduce the number of wavefunction response contributions that
must be considered. The Lagrangian formalism involves the replacement of the
observable $E_{\Theta} (\hat H, \{ \theta_{i} \})$ with an equivalent scalar
quantity $\mathcal{L}_{\Theta} (\hat H, \{ \theta_{i} \}, \{ \tilde \theta_{i}
\})$,
\begin{equation}
\mathcal{L}_{\Theta} (\hat H, \{ \theta_{i} \}, \{ \tilde \theta_{i} \})
\equiv
E_{\Theta} (\hat H, \{ \theta_{i} \})
+
\underbrace{
\sum_{i}
\tilde \theta_{i}
f_{i} (\hat H, \{ \theta_{j} \})
}_{0}
\end{equation}
 in such a way that the definitions of the wavefunctions parameters $\{
\theta_{i} \}$ are built into $\mathcal{L}_{\Theta}$ by Lagrange multipliers $\{
\tilde \theta_{i} \}$. I.e., for each set of (generally nonlinear) equations $\{
f_i(\hat H, \{ \theta_j \}) = 0 \}$ determining a class of wavefunction parameters $\{
\theta_{j} \}$, a Lagrange multiplier term $\sum_{i} \tilde \theta_{i} f_{i}
(\hat H, \{ \theta_{j} \})$ is added into the Lagrangian $\mathcal{L}_{\Theta}$.
Making the Lagrangian stationary with respect to the Lagrange multiplier
parameters $\{ \tilde \theta_{i} \}$, e.g., $\total_{\tilde \theta_{i}}
\mathcal{L}_{\Theta} = 0$ provides the definition of the wavefunction
parameters, e.g., $\Rightarrow \{ f_i (\hat H, \{ \theta_j \}) = 0\}$.  Making
the Lagrangian stationary with respect to the intrinsic wavefunction parameters
$\{ \theta_{i} \}$, e.g., $\total_{\theta_{j}} \mathcal{L}_{\Theta} = 0$
determines the values of the Lagrange multipliers $\{ \tilde \theta_{j} \}$ via
the solution of a set of linear equations (the ``response equations''). Once the
response equations have been solved, the total gradient does not require the
explicit determination of the parameter derivatives with respect to $\zeta$,
\[
\diff{E_{\Theta}}{\zeta}
=
\diff{\mathcal{L}_{\Theta}}{\zeta}
=
\pdiff{\mathcal{L}_{\Theta}}{\hat H}
\diff{\hat H}{\zeta}
+
\underbrace{
\pdiff{\mathcal{L}_{\Theta}}{\theta_{i}}
}_{0}
\pdiff{\theta_{i}}{\zeta}
+
\underbrace{
\pdiff{\mathcal{L}_{\Theta}}{\tilde \theta_{i}}
}_{0}
\pdiff{\tilde \theta_{i}}{\zeta}
\]
\begin{equation}
=
\boxed{
\pdiff{\mathcal{L}_{\Theta}}{\hat H}
\diff{\hat H}{\zeta}
}
\end{equation}
Though note that,
\begin{equation}
\pdiff{\mathcal{L}_{\Theta}}{\hat H}
=
\pdiff{E_{\Theta}}{\hat H}
+
\sum_{i}
\tilde \theta_{i}
\pdiff{f_{i}}{\hat H}
\end{equation}
The last term contains the additional ``response'' contributions to the
gradient - these are generally computationally expedient to build once the
Lagrange multipliers $\{ \tilde \theta_i \}$ are determined. Also note that the
second term in the final chain rule expression is written $\total_{\zeta} \hat
H$ rather than $\partial_{\zeta} \hat H$ - this is because $\hat H$ might itself
have implicit parameters in $\zeta$, for which a separate set response equations
will have to be solved and corresponding response contributions accounted for.
This will occur in a key place in the present manuscript - one set of
Lagrangians will be built, fully variationally optimized, and then
differentiated in the quantum part of the MC-VQE algorithm to determine the
quantum gradient in the AIEM monomer basis. Subsequently, a second set of (now
fully classical) Lagrangians will be built, fully variationally optimized, and
then differentiated to determine the contributions of the classical AIEM matrix
elements to the total gradient. This nesting provides a natural separation
between the quantum and classical parts of the MC-VQE+AIEM gradient algorithm,
and should prove to be a general feature of hybrid quantum/classical analytical
derivative methods in other Hamiltonian representations. 

As a final note, we point out that, by convention, the particular derivative
quantity,
\begin{equation}
\hat \Gamma
\equiv
\diff{E_{\Theta}}{\hat H}
=
\pdiff{\mathcal{L}_{\Theta}}{\hat H}
\end{equation}
is referred to as the ``relaxed density matrix,'' while the corresponding
derivative quantity,
\begin{equation}
\hat \Gamma^{0}
\equiv
\pdiff{E_{\Theta}}{\hat H}
\end{equation}
is referred to as the ``unrelaxed density matrix.''

\subsection{Circuit Gradients and Hessians}

A key technical ingredient in the hybrid quantum/classical analytical derivative
methodology developed below is an efficient and robust approach to compute the
analytical derivatives of quantum circuit observable expectation values with
respect to perturbations in the gate angle parameters of the quantum circuit.
Particularly, finite difference approaches are entirely unacceptable due to
extreme precision requirements in the involved observables, which would require
intractable statistical convergence and noise requirements. Here, we exploit
known tomography formulae for the dependency of quantum circuit observable
expectation values on a handful of active gate angles to develop formulae for
the gradient and Hessian that require determination of the observable
expectation value on a stencil of widely-spaced gate angles, with similar
statistical convergence requirements as required for the original observable
expectation values. 

Consider a quantum circuit starting with the reference state $| \vec 0\rangle$,
proceeding through an arbitrary unitary $\hat U$, thence through a parametrized
$\hat R_y$ gate with parameter $\theta$ acting on qubit $C$ (an arbitrary qubit
index), e.g., $\hat R_y^C (\theta) \equiv e^{-i \theta \hat Y_C}$, and finally
through an arbitrary unitary $\hat V$. I.e., sketched in circuit form,
\begin{equation}
\Qcircuit @R=0.1em @C=0.3em @!R {
\lstick{|0_A\rangle}
 & \multigate{3}{\hat U}
 & \qw
 & \multigate{3}{\hat V}
 & \qw \\
\lstick{|0_B\rangle}
 & \ghost{\hat U}
 & \qw
 & \ghost{\hat V}
 & \qw \\
\lstick{|0_C\rangle}
 & \ghost{\hat U}
 & \gate{\hat R_y^{C} (\theta)}
 & \ghost{\hat V}
 & \qw \\
\lstick{|0_D\rangle}
 & \ghost{\hat U}
 & \qw
 & \ghost{\hat V}
 & \qw \\
}
\end{equation}

The expectation value of an observable operator $\hat O$ is,
\begin{equation}
O (\theta)
=
\langle \hat O \rangle
=
\langle \vec 0 |
\hat U^{\dagger}
\hat R_{y}^{C \dagger} (\theta)
\hat V^{\dagger}
\hat O
\hat V
\hat R_{y}^{C} (\theta)
\hat U
| \vec 0 \rangle
\end{equation}
It can be shown that this expectation value always exactly follows a
sinusoidal dependence,
\begin{equation}
\label{eq:tomography-1}
O (\theta - \theta_0)
=
A 
+
B \cos (2 (\theta - \theta_0))
+
C \sin(2 (\theta - \theta_0))
\end{equation}
where $\theta_0$ is the current setting of the gate angle.  The three
circuit-specific tomography coefficients $A$, $B$, and $C$ can be determined by
sampling $O$ at any three angles. Particularly, sampling at $O^{0} \equiv
O(\theta_0)$ and $O^{\pm} \equiv O(\theta_0 \pm \pi / 4)$ yields,
\begin{equation}
C = (O^{+} - O^{-}) / 2
\end{equation}
\begin{equation}
A = (O^{+} + O^{-}) / 2
\end{equation}
\begin{equation}
B = O^{0} - (O^{+} + O^{-}) / 2
\end{equation}
With the tomography coefficients resolved, the first derivative is now
analytical,
\[
\pdiff{O (\theta)}{\theta}
=
\langle \vec 0 |
\hat U^{\dagger}
\hat R_{y}^{C \dagger} (\theta + \pi / 4)
\hat V^{\dagger}
\hat O
\hat V
\hat R_{y}^{C} (\theta + \pi / 4)
\hat U
| \vec 0 \rangle
\]
\[
-
\langle \vec 0 |
\hat U^{\dagger}
\hat R_{y}^{C \dagger} (\theta - \pi / 4)
\hat V^{\dagger}
\hat O
\hat V
\hat R_{y}^{C} (\theta - \pi / 4)
\hat U
| \vec 0 \rangle
\]
\begin{equation}
\boxed{
=
O (\theta + \pi / 4)
-
O (\theta - \pi / 4)
}
\end{equation}
This is remarkable when we consider the symmetric finite-difference formula with
stepsize $h$,
\[
\pdiff{O (\theta)}{\theta}
=
O (\theta + \pi / 4)
-
O (\theta - \pi / 4)
\]
\begin{equation}
\approx
\frac{1}{2h}
\left [
O (\theta + h)
-
O (\theta - h)
\right ]
\end{equation}
This can easily be extended to higher-order derivatives, by inspection,
\[
\pdiff{^2 O (\theta)}{\theta^2}
=
O (\theta + \pi / 2)
-
2
O (\theta)
+
O (\theta - \pi / 2)
\]
\begin{equation}
\approx
\frac{1}{4h^2}
\left [
O (\theta + 2h)
-
2
O (\theta)
+
O (\theta - 2h)
\right ]
\end{equation}
and,
\[
\pdiff{^2 O (\theta, \theta')}{\theta \partial \theta'}
=
O (\theta + \pi / 4, \theta' + \pi / 4)
-
O (\theta + \pi / 4, \theta' - \pi / 4)
\]
\[
-
O (\theta - \pi / 4, \theta' + \pi / 4)
+
O (\theta - \pi / 4, \theta' - \pi / 4)
\]
\[
\approx
\frac{1}{4 h^2}
\left [
O (\theta + h, \theta' + h)
-
O (\theta + h, \theta' - h)
\right .
\]
\begin{equation}
\left .
-
O (\theta - h, \theta' + h)
+
O (\theta - h, \theta' - h)
\right ]
\end{equation}
A key observation is that the coefficients of the observables in
tomography-based formulae for the gradients and Hessians are of order unity, in
contrast to the $1/h$ or $1/h^2$ coefficients of the finite-difference formulae.
This implies that the tomography-based formulae do not experience subtractive
cancellation, and may be evaluated with similar statistical precision as the
underlying observables to obtain similar absolute accuracy in the derivative
quantities.

Note that many other three-point quadratures can be used to analytically/exactly
resolve the tomography coefficients of Equation \ref{eq:tomography-1} and it's
multi-$\hat R_y$-gate counterpart, and to compute the corresponding derivatives.
For instance, the three-point Fourier grid with collocation points $\{ -\pi / 3,
0, + \pi / 3 \}$ can be used, and only the minutia of the quadrature
collocation-to-tomography coefficient transformation changes.

\section{Explicit MC-VQE+AIEM Energy and Gradient Recipe}

In this section, we carefully define and then differentiate the MC-VQE+AIEM
adiabatic state energies. The flow of this section roughly follows reverse
accumulation automatic differentiation, with a Lagrangian approach used for
implicit portions of the derivatives. 

\subsubsection{Indices}

The following index classes are used in this work,
\begin{itemize}
\item $A$ - Monomer.
\item $I$ - Contracted reference state (CRS) configuration, e.g., configuration
interaction singles (CIS) configuration.
\item $\Xi$ - CRS eigenstate.
\item $\Theta$ - MC-VQE eigenstate.
\item $M$ - CIS quantum circuit angle. 
\item $g$ - VQE entangler quantum circuit angle.
\item $\zeta$ - Nuclear gradient perturbation.
\end{itemize}
Primes are used to distinguish repeated indices.

\subsection{Classical AIEM Energy Stage}

\subsubsection{Monomer Properties}

To begin, for an \emph{ab initio} exciton model (AIEM) with $N$ neutral
monomers, each with two relevant electronic states, and with restricted two-body
interactions computed in the dipole-dipole approximation, compute the following
quantities at the current nuclear positions $\{ \vec r_\zeta \}$.
\begin{itemize}
\item $\epsilon_{\mathrm{H}}^{A}$: The energy of the singlet ground (hole) state of the
monomer.
\item $\epsilon_{\mathrm{P}}^{A}$: The energy of the first singlet excited (particle)
state of the monomer.
\item $\vec \mu_{\mathrm{H}}^{A}$: The total dipole moment of the hole state of the
monomer.
\item $\vec \mu_{\mathrm{P}}^{A}$: the total dipole moment of the particle state of the
monomer.
\item $\vec \mu_{\mathrm{T}}^{A}$: the transition dipole moment between the hole and
particle states.
\item $\vec r_{0}^{A}$: The geometric centroid of the monomer, e.g., the center
of mass of the nuclei, $\vec r_{0}^{A} \equiv \sum_{\zeta} m_{\zeta} \vec
r_{\zeta} / \sum_{\zeta'} m_{\zeta'}$, where $\vec r_{\zeta}$ is the Cartesian
coordinates of atom $\zeta$ and $m_{\zeta}$ is the mass of atom $\zeta$.
\end{itemize}
These quantities can all be computed efficiently by classical electronic
structure methods that scale only with the monomer size, and which are
independent of the total size of the system $N$. For instance, today we will use
ground-state Kohn-Sham density function theory (KS-DFT) to compute the
ground-state monomer properties $\epsilon_{\mathrm{H}}^{A}$ and $\vec
\mu_{\mathrm{H}}^{A}$, and
the Tamm-Dancoff approximation for time-dependent density functional theory
(TDA-TD-DFT) TDA-TD-DFT to compute the excited state and transition properties
$\epsilon_{\mathrm{P}}^{A}$, $\vec \mu_{\mathrm{P}}^{A}$, and $\vec
\mu_{\mathrm{T}}^{A}$ (excited state
and transition dipole moments computed in the unrelaxed expectation value
formulation), though many other electronic structure methods could also be used. 

It may be convenient to also compute the analytical nuclear gradients of all of
these monomer properties at this stage in the computation:
$\epsilon_{\mathrm{H}}^{A,\zeta}$,
$\epsilon_{\mathrm{P}}^{A,\zeta}$, 
$\vec \mu_{\mathrm{H}}^{A,\zeta}$,
$\vec \mu_{\mathrm{P}}^{A,\zeta}$,
$\vec \mu_{\mathrm{T}}^{A,\zeta}$,
and $\vec r_{0}^{A,\zeta}$ will all be required at the very end of this
procedure.  Here, e.g., $\epsilon_{\mathrm{H}}^{A,\zeta} \equiv \total_{\vec
r_\zeta} \epsilon_{\mathrm{H}}^{A}$. The computation of these analytical
derivatives of classical monomer properties often requires a Lagrangian
formalism and the solution of classical response equations, e.g.,
coupled-perturbed Kohn-Sham (CP-KS) equations must be solved in the computation
of $\epsilon_{\mathrm{P}}^{A,\zeta}$, $\vec \mu_{\mathrm{H}}^{A,\zeta}$, $\vec
\mu_{\mathrm{P}}^{A,\zeta}$, and $\vec \mu_{\mathrm{T}}^{A,\zeta}$. Note that
these response equations may be solved separately from the quantum response
equations that will appear later in the procedure. Alternatively, if one wishes
to follow the spirit of the Lagrangian formalism to its zenith, these monomer
derivative property computations can be deferred to the very end of the
procedure, in which case they will appear as chain rule terms contracted against
density matrix quantities from the quantum portion of the algorithm. This
formulation can provide enhanced screening if some density matrix elements are
small, and can also reduce the number of classical response equations that must
be explicitly computed. E.g., the hole dipole chain-rule contribution is
$E_{\Theta}^{A,\zeta} \leftarrow \vec \eta_{\mathrm{H}}^{A,\Theta} \vec
\mu_{\mathrm{H}}^{A,\zeta}$, which can be computed with a single CP-KS response
if the hole dipole density matrix $\vec \eta_{\mathrm{H},\Theta}^{A}$ is
available, vs three separate CP-KS response contributions to compute the
Cartesian hole dipole derivatives $\vec \mu_{\mathrm{H}}^{A,\zeta}$ beforehand.

\subsubsection{Dimer Interaction Matrix Elements}

In the present work, electrostatic interactions between monomers are computed in
the dipole-dipole approximation,
\begin{equation}
\label{eq:v_dipole_dipole}
v^{AA'}
=
\frac{\vec \mu_{A} \cdot \vec \mu_{A'}}{r_{AA'}^3} 
- 3 
\frac{(\vec \mu_{A} \cdot \vec r_{AA'}) (\vec
\mu_{A'} \cdot \vec r_{AA'})}{r_{AA'}^5}
\end{equation}
Here $\vec r_{AA'} \equiv \vec r_{0}^{A'} - \vec r_{0}^{A}$, and $r_{AA'} = \sqrt{|
\vec r_{AA'}|^2}$. The dipole moments run over the types of $\mathrm{H}$,
$\mathrm{T}$, and $\mathrm{P}$ for monomers $A$ and $A'$, leading to $9\times$
possible two-body coupling matrix elements for a given $A,A'$ pair: 
$v_{\mathrm{HH}}^{AA'}$,
$v_{\mathrm{HT}}^{AA'}$,
$v_{\mathrm{HP}}^{AA'}$,
$v_{\mathrm{TH}}^{AA'}$,
$v_{\mathrm{TT}}^{AA'}$,
$v_{\mathrm{TP}}^{AA'}$,
$v_{\mathrm{PH}}^{AA'}$,
$v_{\mathrm{PT}}^{AA'}$, and
$v_{\mathrm{PP}}^{AA'}$. Note the symmetries,
$v_{\mathrm{HH}}^{AA'} = v_{\mathrm{HH}}^{A'A}$,
$v_{\mathrm{TT}}^{AA'} = v_{\mathrm{TT}}^{A'A}$,
$v_{\mathrm{PP}}^{AA'} = v_{\mathrm{PP}}^{A'A}$,
$v_{\mathrm{HT}}^{AA'} = v_{\mathrm{TH}}^{A'A}$,
$v_{\mathrm{HP}}^{AA'} = v_{\mathrm{PH}}^{A'A}$, and
$v_{\mathrm{TP}}^{AA'} = v_{\mathrm{PT}}^{A'A}$.
Due to the $r_{AA'}^{-3}$ decay of these matrix elements, distant $A,A'$ pairs may
often be eliminated with insignificant errors (e.g., today we assert a
nearest-neighbor-only coupling model, for simplicity). We denote the set of
significant $A,A'$ pairs included in the AIEM by the notation $<A,A'>$ (including
both $A>A'$ and $A<A'$). Note that $A=A'$ pairs are never included - these are
already accounted for in the one-body matrix elements
$\epsilon_{\mathrm{H}}^{A}$ and $\epsilon_{\mathrm{P}}^{A}$.

Also note that it is entirely possible to go beyond the dipole-dipole
approximation for these dimer interaction matrix elements: in prior work we have
also used fully \emph{ab initio} matrix elements involving the electrostatic
interactions between state or transition densities to account for higher-order
multipole contributions and charge penetration terms. Such matrix elements could
easily be used within MC-VQE+AIEM energies and derivatives, though the full
Lagrangian formalism discussed above would be useful to avoid having to
explicitly form the derivatives of the two-body matrix elements $v^{AA',\zeta}$.

\subsubsection{Monomer-Basis AIEM Hamiltonian}

In the monomer basis, the AIEM Hamiltonian is succinctly written as,
\begin{equation}
\hat H
\equiv
\hat H^{(1)}
+
\hat H^{(2)}
\end{equation}
\[
=
\sum_{A}
\sum_{p,q \in [0,1]}
(p_A | \hat h | q_A) 
|p_A \rangle \langle q_A |
\]
\[
+
\frac{1}{2}
\sum_{<A,A'>}
\sum_{p,q,r,s \in [0,1]}
(p_A q_A | \hat h | r_{A'} s_{A'}) 
|p_A \rangle \langle q_A |
\otimes
|r_{A'} \rangle \langle s_{A'} |
\]
Where now $|0_A\rangle$ refers to the monomer ground state (hole), $|1_A\rangle$
refers to the monomer excited state (particle), and in chemist's notation $|00)$
refers to H (hole), $|11)$ refers to P (particle), and $|01)$ or $|10)$ refers
to T (transition). Note that the correspondence of the matrix elements is,
\begin{equation}
(0_A | \hat h | 0_A) = \epsilon_{\mathrm{H}}^{A}
\end{equation}
\begin{equation}
(1_A | \hat h | 1_A) = \epsilon_{\mathrm{P}}^{A}
\end{equation}
\begin{equation}
(0_A | \hat h | 1_A) = 
(1_A | \hat h | 0_A) = 0
\end{equation}
\begin{equation}
(0_A 0_A | \hat v | 0_{A'} 0_{A'}) = v_{\mathrm{HH}}^{A{A'}}
\end{equation}
\begin{equation}
(0_A 0_A | \hat v | 1_{A'} 1_{A'}) = v_{\mathrm{HP}}^{A{A'}}
\end{equation}
\begin{equation}
(1_A 1_A | \hat v | 0_{A'} 0_{A'}) = v_{\mathrm{PH}}^{A{A'}}
\end{equation}
\begin{equation}
(1_A 1_A | \hat v | 1_{A'} 1_{A'}) = v_{\mathrm{PP}}^{A{A'}}
\end{equation}
\begin{equation}
(0_A 0_A | \hat v | 0_{A'} 1_{A'}) = 
(0_A 0_A | \hat v | 1_{A'} 0_{A'}) = 
v_{\mathrm{HT}}^{A{A'}}
\end{equation}
\begin{equation}
(0_A 1_A | \hat v | 0_{A'} 0_{A'}) = 
(1_A 0_A | \hat v | 0_{A'} 0_{A'}) = 
v_{\mathrm{TH}}^{A{A'}}
\end{equation}
\begin{equation}
(1_A 1_A | \hat v | 0_{A'} 1_{A'}) = 
(1_A 1_A | \hat v | 1_{A'} 0_{A'}) = 
v_{\mathrm{PT}}^{A{A'}}
\end{equation}
\begin{equation}
(0_A 1_A | \hat v | 1_{A'} 1_{A'}) = 
(1_A 0_A | \hat v | 1_{A'} 1_{A'}) = 
v_{\mathrm{TP}}^{A{A'}}
\end{equation}
\begin{equation}
(0_A 1_A | \hat v | 0_{A'} 1_{A'}) = 
(0_A 1_A | \hat v | 1_{A'} 0_{A'}) 
\end{equation}
\[
=
(1_A 0_A | \hat v | 0_{A'} 1_{A'}) = 
(1_A 0_A | \hat v | 1_{A'} 0_{A'}) = 
v_{\mathrm{TT}}^{A{A'}}
\]
In practice, the monomer-basis representation of the AIEM Hamiltonian is only
used for a formal definition: in computational practice, we transition
immediately from the definition of the matrix elements in the previous section
to the Pauli-basis representation of the AIEM Hamiltonian in the following
section.

\subsubsection{Pauli-Basis AIEM Hamiltonian}

After some straightforward algebra, the AIEM Hamiltonian can be succinctly
re-written in terms of Pauli operators,
\[
\hat H
\equiv
\mathcal{E}
+
\mathcal{H}^{(1)}
+
\mathcal{H}^{(2)}
=
\mathcal{E} \hat I
+
\sum_{A}
\mathcal{Z}_{A}
\hat Z_{A}
+
\mathcal{X}_{A}
\hat X_{A}
\]
\[
+
\frac{1}{2}
\sum_{<A,A'>}
\mathcal{XX}_{AA'}
\hat X_{A} \otimes \hat X_{A'}
+
\mathcal{XZ}_{AA'}
\hat X_{A} \otimes \hat Z_{A'}
\]
\begin{equation}
\label{eq:aiem_pauli_H}
+
\mathcal{ZX}_{AA'}
\hat Z_{A} \otimes \hat X_{A'}
+
\mathcal{ZZ}_{AA'}
\hat Z_{A} \otimes \hat Z_{A'}
\end{equation}
The matrix elements are,
\[
\mathcal{E}
=
\sum_{A}
(
\epsilon_{\mathrm{H}}^{A}
+
\epsilon_{\mathrm{P}}^{A}
) / 2
+
\frac{1}{2}
\sum_{<A,A'>}
(
v_{\mathrm{HH}}^{AA'}
+
v_{\mathrm{HP}}^{AA'}
+
v_{\mathrm{PH}}^{AA'}
+
v_{\mathrm{PP}}^{AA'}
) / 4
\]
\begin{equation}
\label{eq:aiem_pauli_E}
\equiv
\sum_{A}
\epsilon_{\mathrm{S}}^{A}
+
\frac{1}{2}
\sum_{<A,A'>}
v_{\mathrm{SS}}^{AA'}
\end{equation}
\[
\mathcal{X}_{A}
=
\underbrace{
\epsilon_{\mathrm{T}}^{A}
}_{0}
+
\frac{1}{2}
\sum_{A'}
(
v_{\mathrm{TH}}^{AA'}
+
v_{\mathrm{TP}}^{AA'}
) / 2
+
(
v_{\mathrm{HT}}^{A'A}
+
v_{\mathrm{PT}}^{A'A}
) / 2
\]
\begin{equation}
\equiv
\underbrace{
\epsilon_{\mathrm{T}}^{A}
}_{0}
+
\frac{1}{2}
\sum_{A'}
v_{\mathrm{TS}}^{AA'}
+
v_{\mathrm{ST}}^{A'A}
\end{equation}
\[
\mathcal{Z}_{A}
=
(
\epsilon_{\mathrm{H}}^{A}
-
\epsilon_{\mathrm{P}}^{A}
) / 2
\]
\[
+
\frac{1}{2}
\sum_{A'}
(
v_{\mathrm{HH}}^{AA'}
+
v_{\mathrm{HP}}^{AA'}
-
v_{\mathrm{PH}}^{AA'}
-
v_{\mathrm{PP}}^{AA'}
) / 4
\]
\[
+
(
v_{\mathrm{HH}}^{A'A}
+
v_{\mathrm{PH}}^{A'A}
-
v_{\mathrm{HP}}^{A'A}
-
v_{\mathrm{PP}}^{A'A}
) / 4
\]
\begin{equation}
\equiv
\epsilon_{\mathrm{D}}^{A}
+
\frac{1}{2}
\sum_{A'}
v_{\mathrm{SD}}^{AA'}
+
v_{\mathrm{DS}}^{A'A}
\end{equation}
\begin{equation}
\mathcal{XX}_{AA'}
=
v_{\mathrm{TT}}^{AA'}
\end{equation}
\begin{equation}
\mathcal{XZ}_{AA'}
=
(
v_{\mathrm{TH}}^{AA'}
-
v_{\mathrm{TP}}^{AA'}
) / 2
\equiv
v_{\mathrm{TD}}^{AA'}
\end{equation}
\begin{equation}
\mathcal{ZX}_{AA'}
=
(
v_{\mathrm{HT}}^{AA'}
-
v_{\mathrm{PT}}^{AA'}
) / 2
\equiv
v_{\mathrm{DT}}^{AA'}
\end{equation}
\begin{equation}
\label{eq:aiem_pauli_ZZ}
\mathcal{ZZ}_{AA'}
=
(
v_{\mathrm{HH}}^{AA'}
-
v_{\mathrm{HP}}^{AA'}
-
v_{\mathrm{PH}}^{AA'}
+
v_{\mathrm{PP}}^{AA'}
) / 4
\equiv
v_{\mathrm{DD}}^{AA'}
\end{equation}
Here the new letters are S (sum) and D (difference). $\hat I$ operators
correspond to S, $\hat X$ operators correspond to T, and $\hat Z$ operators
correspond to D.

\subsection{Quantum MC-VQE Energy Stage}

\subsubsection{Formal Definition: MC-VQE Eigenstates}

MC-VQE approximates the exact diagonalization of the Pauli Hamiltonian $\hat H$
by producing a number $N_{\Theta}$ of MC-VQE approximate eigenstates $ |
\Psi_{\Theta} \rangle $.
\begin{equation}
| \Psi_{\Theta} \rangle
\equiv
\hat U ( \theta_{g}  )
\sum_{\Xi}
| \Phi_{\Xi} \rangle
V_{\Xi \Theta}
\end{equation}
\begin{equation}
=
\hat U ( \theta_{g}  )
\sum_{I}
\sum_{\Xi}
| I \rangle
C_{I\Xi}
V_{\Xi \Theta}
\end{equation}
\begin{equation}
=
\hat U ( \theta_{g}  )
\sum_{I}
| I \rangle
\Gamma_{I\Theta}
\end{equation}
\begin{equation}
=
\hat U ( \theta_{g}  )
| \Gamma_{\Theta} \rangle
\end{equation}
The MC-VQE eigenstates are orthonormal,
\begin{equation}
\langle \Psi_{\Theta} | \Psi_{\Theta'} \rangle 
= 
\delta_{\Theta \Theta'}
\end{equation}
and within the MC-VQE contracted, entangled subspace, the Hamiltonian is
diagonal,
\begin{equation}
\langle \Psi_{\Theta} | \hat H | \Psi_{\Theta'} \rangle 
=
E_{\Theta}
\delta_{\Theta \Theta'}
\end{equation}

Specific quantities are defined below.

\subsubsection{Configuration Interaction Singles Contracted Reference States}

To begin, we classically determine a number $N_{\Xi} = N_{\Theta}$ of
``contracted reference states'' (CRS), $\{ | \Phi_{\Xi} \rangle \}$, by solving
a polynomial-scaling electronic structure problem to ``sketch out'' the shapes
of the desired electronic states. One particularly appealing choice for the
contracted reference states is the set of eigenstates of a restricted
configuration interaction method, such as configuration interaction singles
(CIS),

Form the configuration interaction singles (CIS) Hamiltonian. This matrix is
indexed by a restricted set of configurations $\{ |I\rangle \} \equiv \{ | 0
\rangle \} + \{ | A \rangle \}$ consisting of the reference configuration
$|0\rangle \equiv |00\ldots0\rangle$, followed by the $N$ singly-excited
configurations $\{ | A \rangle \}$ including $|10\ldots0\rangle$,
$|01\ldots0\rangle$, and finally $|00\ldots1\rangle$ ($N+1$ configurations
total). The CIS Hamiltonian has the block form,
\begin{equation}
\hat H_{\mathrm{CIS}}
\equiv
\left [
\begin{array}{c|c}
H_{00} & H_{0A'} \\
\hline
H_{A0} & H_{AA'} \\
\end{array}
\right ]
\end{equation}
The reference-reference block is,
\begin{equation}
\label{eq:cis_H_00}
H_{00}
=
\langle 0 | \hat H | 0 \rangle
=
\mathcal{E}
+
\sum_{A}
\mathcal{Z_{A}}
+
\frac{1}{2}
\sum_{<A, A'>}
\mathcal{ZZ_{AA'}}
\equiv
E_{\mathrm{ref}}
\end{equation}
The diagonal singles-singles block is,
\begin{equation}
H_{AA}
=
\langle A | \hat H | A \rangle
=
E_{\mathrm{ref}}
-
2 \mathcal{Z}_{A}
-
\sum_{A'} 
\left [
\mathcal{ZZ}_{AA'}
+
\mathcal{ZZ}_{A'A}
\right ]
\end{equation}
The reference-singles block is,
\begin{equation}
H_{0A}
=
H_{A0}
=
\langle 0 | \hat H | A \rangle
=
\mathcal{X}_{A}
+
\frac{1}{2}
\sum_{A'} 
\left [
\mathcal{XZ}_{AA'}
+
\mathcal{ZX}_{A'A}
\right ]
\end{equation}
The off-diagonal singles-singles block is,
\begin{equation}
\label{eq:cis_H_CD}
H_{A\neq A'}
=
\langle A | \hat H | A' \rangle|_{A\neq A'}
=
\mathcal{XX}_{AA'}
\end{equation}

Now, classically diagonalize the CIS Hamiltonian to obtain the CIS contracted
reference states (CRS),
\begin{equation}
|\Phi_{\Xi} \rangle
\equiv
\sum_{I}
| I \rangle
C_{I\Xi}
: \
\sum_{J}
H_{I J}
C_{J\Xi}
=
C_{I\Xi}
E_{\Xi}^{\mathrm{CIS}}
,
\end{equation}
\[
\sum_{I}
C_{I\Xi}
C_{I\Xi'}
=
\delta_{\Xi \Xi'}
\]
Note that we often focus on $N_{\Theta} = N_{\Xi} \leq N_{I}$ - from this point
forward, it is assumed that any $\Theta$ or $\Xi$ index only runs up to
$N_{\Theta}$.

Note that other choices of contracted reference states would lead to different
forms/results for the MC-VQE eigenstates and different response contributions in
analytical derivatives, but the manipulations would be similar. One particularly
interesting alternative example is the use of single configurations (e.g.,
``determinants'') or \emph{a priori} known combinations of several
configurations (e.g., ``configuration state functions'')
- here, no implicit equations are solved to determine the contracted reference
  states, and so no response terms would appear in the corresponding analytical
derivatives.

\subsubsection{Technical Detail: CIS State Preparation Circuit}

A quantum circuit to prepare the CIS state $| \Phi \rangle \equiv \sum_{I} |I
\rangle C_{I} \equiv \hat U_{\mathrm{CIS}} | 0 \rangle$ is sketched for $N=4$,
\begin{widetext}
\begin{equation}
\label{eq:cis_circuit}
\Qcircuit @R=0.5em @C=0.5em {
\lstick{|0_A\rangle}
 & \gate{R_{y} (\theta_0)}
 & \ctrl{1}
 & \qw
 & \targ
 & \qw
 & \qw
 & \qw
 & \qw
 & \qw
 & \qw
 & \qw \\
\lstick{|0_B\rangle}
 & \gate{R_{y} (- \theta_{AB} / 2)}
 & \ctrl{-1}
 & \gate{R_{y} (+ \theta_{AB} / 2)}
 & \ctrl{-1}
 & \ctrl{1}
 & \qw
 & \targ
 & \qw
 & \qw
 & \qw
 & \qw \\
\lstick{|0_C\rangle}
 & \qw
 & \qw
 & \qw
 & \gate{R_{y} (- \theta_{BC} / 2)}
 & \ctrl{-1}
 & \gate{R_{y} (+ \theta_{BC} / 2)}
 & \ctrl{-1}
 & \ctrl{1}
 & \qw
 & \targ
 & \qw \\
\lstick{|0_D\rangle}
 & \qw
 & \qw
 & \qw
 & \qw
 & \qw
 & \qw
 & \gate{R_{y} (- \theta_{CD} / 2)}
 & \ctrl{-1}
 & \gate{R_{y} (+ \theta_{CD} / 2)}
 & \ctrl{-1}
 & \qw \\
}
\end{equation}
\end{widetext}
The $R_{y} (\theta_0)$ gate controls the amplitude of the reference ket
$|00\ldots\rangle$, while the composite two-body gates between each pair of
wires control the amplitudes of each singly-excited ket $|10\ldots\rangle$,
$|01\ldots\rangle$, etc. Specifically, the composite two-body gate is,
\begin{widetext}
\begin{equation}
\label{eq:Fy_circuit}
\Qcircuit @R=1.0em @C=1.0em {
& \qw & \ctrl{1} & \qw & \targ & \qw  \\
& \gate{R_{y} (-\theta/2)} & \ctrl{-1} & \gate{R_{y} (+\theta/2)} & \ctrl{-1} & \qw \\
}
=
\left [
\begin{array}{cccc}
 1 & & & \\
 & & +s(\theta) & -c(\theta) \\
 & & +c(\theta) & +s(\theta) \\
 & 1 & & \\
\end{array}
\right ]
\end{equation}
\end{widetext}
The action of this composite two-body gate on a statevector of the form $\mu |00
\rangle + \mu' | 10 \rangle$ yields a statevector of the form $\mu | 00 \rangle
+ \alpha | 10 \rangle + \beta | 01 \rangle$, where $\alpha = \mu' \cos(\theta)$
and $\beta = \mu' \sin(\theta)$.
\begin{equation}
\left [
\begin{array}{cccc}
 1 & & & \\
 & & +s(\theta) & -c(\theta) \\
 & & +c(\theta) & +s(\theta) \\
 & 1 & & \\
\end{array}
\right ]
\left [
\begin{array}{c}
 \mu \\
 0 \\
 \mu' \\
 0 \\
\end{array}
\right ]
=
\left [
\begin{array}{c}
 \mu \\
 \mu' s (\theta) \\
 \mu' c (\theta) \\
 0 \\
\end{array}
\right ]
\end{equation}
E.g., this gate transfers single excitation from qubit $A$ to qubit $B$.  We
note that there are many possibly choices for the composite two-body gates and
the overall CIS state preparation circuit; the definition used here is merely
pragmatic.  Overall, our selected CIS state preparation circuit requires a
linear number of gates and, as written, requires linear depth with only linear
nearest-neighbor qubit connectivity - a drastic simplification over our previous
efforts which used a quadratic sequence of CNOT gates. It is worth noting that
it appears that this state-preparation circuit can be extended to use
logarithmic depth at the cost of higher two-qubit connectivity and modified
definitions of the gate angles, e.g., by moving from a linear control sequence
to a binary-tree control sequence.

By inspection, it is straightforward to classically determine the CIS circuit
gate angles from the CIS wavefunction amplitudes,
\begin{equation}
\theta_{0}
=
\cos^{-1} \left (C_{0} / \sqrt{C_{0}^2 + C_{A}^{2} + \ldots + C_{
N}^2} \right )
\end{equation}
\begin{equation}
\theta_{AB}
=
\cos^{-1} \left (C_{A} / \sqrt{C_{A}^2 + C_{B}^{2} + \ldots + C_{
N}^2} \right )
\end{equation}
\begin{equation}
\vdots
\end{equation}
\begin{equation}
\theta_{MN}
=
\mathrm{sign} (C_{N})
\cos^{-1} \left (C_{M} / \sqrt{C_{M}^2 + C_{N}^2} \right )
\end{equation}
Note that the raw value of the last CIS coefficient $C_N$ is not used inside of
the arccos formula set, so an additional phase factor appears in the last angle
to preserve the total information content of the CIS coefficient vector.

A succinct classical function to return the $N$ CIS circuit angles $\{ \theta_M
\}$ for an arbitrary normalized CIS state with $N+1$ coefficients $\{ C_{I} \}$
is,
\begin{equation}
\theta_{M} [ C_{I} ]
\equiv
P_{M}
\cos^{-1}
\left (
\frac{
C_{M}
}{
\sqrt{
\sum_{L=M}^{N}
C_{L}^2
}
}
\right )
, \ M \in [0, N-1]
\end{equation}
\begin{equation}
P_{M} \equiv 
\left \{
\begin{array}{ll}
\mathrm{sign} (C_{N}) & M = N - 1 \\
1 & \mathrm{else} \\
\end{array}
\right .
\end{equation}
We will need this function in several places, with a number of different choices
for the $N+1$ input CIS coefficients $\{ C_{I} \}$. Another function that will
prove to be extremely useful is the contraction of an arbitrary $N$-dimensional
vector $d_M$ with the Jacobian $\partial\theta_M [C_{I'}] / \partial
C_{I}$,
\begin{equation}
\label{eq:cis_jacobian}
\diff{O}{C_{I}}
[
C_{I'},
d_M
]
\equiv
\sum_{M}
\underbrace{
\pdiff{O}{\theta_{M} [C_{I'}]}
}_{d_{M}}
\pdiff{\theta_{M} [C_{I'}]}{C_{I}}
\end{equation}
\[
=
\sum_{M}
d_{M}
P_{M}
\left [
-
\frac{\delta_{MI}}{
\sqrt{
\sum_{L=M+1}^{N}
C_{L}^2
}
}
\right .
\]
\[
\left .
+
\frac{C_{M} C_{I} \delta_{I \geq M}}
{
\sqrt{
\sum_{L=M+1}^{N}
C_{L}^2
}
\left (
\sum_{K=M}^{N}
C_{K}^2
\right )
}
\right ]
\]

\subsubsection{CIS Circuit Angles}

Using the recipe from the previous section, compute the $N$ CIS circuit angles
from the $N+1$ CIS coefficients, for each state $\Xi$:
\begin{equation}
\theta_{M} [C_{I \Xi}]
\end{equation}

\subsubsection{State-Averaged VQE}

A key step in MC-VQE is the use of a state-averaged VQE entangler operator $\hat
U_{\mathrm{VQE}} (\theta_{g})$ to maximally decouple the contracted reference states $\{
| \Phi_{\Xi} \rangle \}$ from the rest of the Hilbert space, i.e., using the VQE
entangler operator to maximally block-diagonalize the Hamiltonian.  This VQE
entangler operator acts over the full qubit Hilbert space (tractable because of
the use of a quantum computer), and is specified by a polynomial number of
nonlinear parameters $\{ \theta_{g} \}$.  The maximal decoupling condition is
equivalent (in the 2-norm sense) to minimizing the state-averaged energy,
\begin{equation}
\hat U_{\mathrm{VQE}} ( \theta_{g} ) =
\underset{\hat U_{\mathrm{VQE}} ( \theta_{g} )}{\mathrm{argmin}}
[
\bar E
]
\end{equation}
which is defined as,
\begin{equation}
\bar E
\equiv
\frac{1}{N_{\Theta}}
\sum_{\Theta}
E_{\Theta}
\equiv
\frac{1}{N_{\Theta}}
\sum_{\Theta}
\langle \Psi_{\Theta} | \hat H | \Psi_{\Theta} \rangle
\end{equation}
\begin{equation}
=
\frac{1}{N_{\Theta}}
\sum_{\Xi}
\langle \Phi_{\Xi} | \hat U_{\mathrm{VQE}}^\dagger \hat H \hat U_{\mathrm{VQE}} | \Phi_{\Xi} \rangle
\end{equation}
where $\hat H$ is the Hamiltonian operator in the qubit Hilbert space.  The
MC-VQE state energy is defined as,
\begin{equation}
E_{\Theta}
\equiv
\langle \Psi_{\Theta} |
\hat H
| \Psi_{\Theta} \rangle
\end{equation}
The weak form of the state-averaged variational condition is the zero-gradient
condition,
\begin{equation}
\pdiff{\bar E}{\theta_{g}} = 0
\end{equation}

Let us consider a key set of Pauli density matrices, which are defined as a
function for an arbitrary set of CIS state preparation angles $\{ \theta_M \}$
and an arbitrary set of VQE $\hat R_y$ gate parameter angles $\{ \theta_{g} \}$,
e.g.,
\begin{equation}
\lambda_{\mathcal{Z}_{A}} 
[
 \theta_{g} ,
 \theta_{M} 
]
\equiv
\end{equation}
\[
\langle 0 |
\hat U_{\mathrm{CIS}}^{\dagger} [  \theta_{M}  ]
\hat U_{\mathrm{VQE}}^{\dagger} [  \theta_{g}  ]
\hat Z_A
\hat U_{\mathrm{VQE}} [  \theta_{g}  ]
\hat U_{\mathrm{CIS}} [  \theta_{M}  ]
| 0 \rangle
\]
or,
\begin{equation}
\Lambda_{\mathcal{ZZ}_{AA'}} 
[
 \theta_{g} ,
 \theta_{M} 
]
\equiv
\end{equation}
\[
\langle 0 |
\hat U_{\mathrm{CIS}}^{\dagger} [  \theta_{M}  ]
\hat U_{\mathrm{VQE}}^{\dagger} [  \theta_{g}  ]
\hat Z_A
\otimes
\hat Z_{A'}
\hat U_{\mathrm{VQE}} [  \theta_{g}  ]
\hat U_{\mathrm{CIS}} [  \theta_{M}  ]
| 0 \rangle
\]
These can be evaluated by statistically-converged tomography measurements of
quantum circuits passing through a CIS-state preparation stage $\hat
U_{\mathrm{CIS}}$ and a VQE stage $\hat U_{\mathrm{VQE}}$ with the specified
parameters. Many of the measurements correspond to commuting operators in
disjoint sets of qubits, and can be performed in parallel. Measurements in the
$\hat X$ basis can be easily made by postpending the circuit with a Hadamard
gate(s) in the appropriate position(s). Note that we will require the one-body
Pauli density matrices for all one-body elements 
$\mathcal{Z}_{A}$ and 
$\mathcal{X}_{A}$, and for all two-body
elements 
$\mathcal{XX}_{AA'}$,
$\mathcal{XZ}_{AA'}$,
$\mathcal{ZX}_{AA'}$,
$\mathcal{ZZ}_{AA'}$, for all $A,A'$ pairs in $<A,A'>$.

The energy of a VQE-entangled CIS contracted reference state is, with
appropriate parameters,
\begin{equation}
\varepsilon
[
\{ \theta_{g} ,
 \theta_{M} 
]
\equiv
\mathcal{E}
+
\sum_{A}
\lambda_{\mathcal{Z}_{A}}
\mathcal{Z}_{A}
+
\lambda_{\mathcal{X}_{A}}
\mathcal{X}_{A}
\end{equation}
\[
+
\frac{1}{2}
\sum_{<A,A'>}
\Lambda_{\mathcal{XX}_{AA'}}
\mathcal{XX}_{AA'}
+
\Lambda_{\mathcal{XZ}_{AA'}}
\mathcal{XZ}_{AA'}
\]
\[
+
\Lambda_{\mathcal{ZX}_{AA'}}
\mathcal{ZX}_{AA'}
+
\Lambda_{\mathcal{ZZ}_{AA'}}
\mathcal{ZZ}_{AA'}
\]
This energy can be computed classically after the Pauli density matrices have
been evaluated.  Note that only the Pauli density matrices $\{ \lambda \}$ and
$\{\Lambda \}$ depend on the CIS or VQE parameters $\{ \theta_M \}$ or $\{
\theta_{g} \}$. 

A useful specialization to VQE-entangled CIS contracted reference states is,
\begin{equation}
\lambda_{\mathcal{Z}_{A}}^{\Xi} [ \theta_g  ]
\equiv
\lambda_{\mathcal{Z}_{A}}
[
 \theta_{g} ,
 \theta_{M} [  C_{I \Xi}  ] 
]
\end{equation}
and the corresponding energy is,
\begin{equation}
\varepsilon^{\Xi}
[
 \theta_{g} 
]
\equiv
H_{\Xi \Xi}
=
\varepsilon 
[
 \theta_{g} ,
 \theta_{M} [  C_{I \Xi}  ] 
]
\end{equation}

A further specialization to state-averaged VQE-entangled CIS contracted reference
states is,
\begin{equation}
\bar \lambda_{\mathcal{Z}_{A}} [  \theta_g  ]
\equiv
\frac{1}{N_{\Theta}}
\sum_{\Xi}
\lambda_{\mathcal{Z}_{A}}^{\Xi} [  \theta_g ]
\end{equation}
and the corresponding state-averaged MC-VQE energy is,
\begin{equation}
\bar E [ \theta_g  ]
\equiv
\frac{1}{N_{\Theta}}
\sum_{\Xi}
\varepsilon^{\Xi} [  \theta_g  ]
\end{equation}

The goal of the hybrid quantum/classical MC-VQE optimization procedure is to
find the set of VQE parameters $\{ \theta_{g}^{*} \}$ that minimize the
state-averaged VQE energy,
\begin{equation}
\{
\theta_{g}^*
\}
\equiv
\underset{
\{
\theta_{g}
\}
}{
\mathrm{argmin}
}
\left (
\bar E
[
\theta_{g}
]
\right )
\end{equation}
To aid in the optimization, it can be useful to have the gradient of the
state-averaged VQE energy for a given set of VQE parameters,
\begin{equation}
\pdiff{\bar E [ \theta_{g'} ]}{\theta_{g}}
\end{equation}
E.g., the weak form of the state-averaged VQE functional is the stationary
condition,
\begin{equation}
\pdiff{\bar E [ \theta_{g'}^* ]}{\theta_{g}}
= 0 
\ \forall \ g
\end{equation}
Moreover (subject to noise constraints), the state-averaged VQE energy
optimization procedure can use gradient-based optimization algorithms such as
L-BFGS to enhance the convergence of the MC-VQE energy functional.

The gradients with respect to the VQE parameters are,
\begin{equation}
\pdiff{\bar E[ \theta_{g'} ]}{\theta_{g}}
=
\frac{1}{N_{\Theta}}
\sum_{\Xi}
\pdiff{\varepsilon^{\Xi} [ \theta_{g'} ]}{\theta_{g}}
\end{equation}
and,
\begin{equation}
\pdiff{\varepsilon^{\Xi} [ \theta_{g'} ]}{\theta_{g}}
\equiv
\sum_{A}
\left (
\partial_{\theta_g}
\lambda_{\mathcal{Z}_{A}}^{\Xi}
\right )
\mathcal{Z}_{A}
+
\left (
\partial_{\theta_g}
\lambda_{\mathcal{X}_{A}}^{\Xi}
\right )
\mathcal{X}_{A}
\end{equation}
\[
+
\frac{1}{2}
\sum_{<A,A'>}
\left (
\partial_{\theta_g}
\Lambda_{\mathcal{XX}_{AA'}}^{\Xi}
\right )
\mathcal{XX}_{AA'}
+
\left (
\partial_{\theta_g}
\Lambda_{\mathcal{XZ}_{AA'}}^{\Xi}
\right )
\mathcal{XZ}_{AA'}
\]
\[
+
\left (
\partial_{\theta_g}
\Lambda_{\mathcal{ZX}_{AA'}}^{\Xi}
\right )
\mathcal{ZX}_{AA'}
+
\left (
\partial_{\theta_g}
\Lambda_{\mathcal{ZZ}_{AA'}}^{\Xi}
\right )
\mathcal{ZZ}_{AA'}
\]
and, by the results from analytical circuit gradients,
\begin{equation}
\partial_{\theta_{g}}
\lambda_{\mathcal{Z}_{A}}^{\Xi}
=
\lambda_{\mathcal{Z}_{A}}^{\Xi,\theta_g+\pi/4}
-
\lambda_{\mathcal{Z}_{A}}^{\Xi,\theta_g-\pi/4}
\end{equation}
Here $
\lambda_{\mathcal{Z}_{A}}^{\Xi,\theta_g+\pi/4}
$ is a shorthand to indicate that the Pauli matrix should be reevaluated with all
parameters frozen, but with $\theta_g$ updated to $\theta_g + \pi / 4$.

The sole goal/directive of this subsubsection is to determine the optimal
state-averaged VQE parameters $\{ \theta_{g}^* \}$. The diagonal subspace
Hamiltonian matrix elements $H_{\Xi \Xi} \equiv \varepsilon^{\Xi}$ are
determined as a side externality of this computation.

\subsubsection{Technical Detail: VQE Entangler Parametrization} 

The definition and parametrization of the VQE entangler circuit is something of
an art.  In any configuration-space basis, the adiabatic eigenfunctions of the
real electronic or \emph{ab initio} exciton Hamiltonian can be written as real,
orthonormal vectors with arbitrary total phase of $\pm 1$. Therefore, the VQE
entangler operator $\hat U$ can be restricted to $SO(2^N)$ without loss.  We
have previously\cite{Parrish:2019:230401} elected to construct the total VQE
entangler circuit for the \emph{ab initio} exciton model by placing a two-body
entangler restricted to $SO(4)$ at each two-body interaction site in the
exciton Hamiltonian. E.g., for a linear arrangement,
\begin{widetext}
\begin{equation}
\Qcircuit @R=1.0em @C=1.0em {
\lstick{| A \rangle}
& \multigate{1}{U_{2}^{AB}}
& \qw
& \qw \\
\lstick{| B \rangle}
& \ghost{U_{2}^{AB}}
& \multigate{1}{U_{2}^{BN}}
& \qw \\
\lstick{| N \rangle}
& \multigate{1}{U_{2}^{NZ}}
& \ghost{U_{2}^{BN}}
& \qw \\
\lstick{| Z \rangle}
& \ghost{U_{2}^{NZ}}
& \qw
& \qw \\
}
\end{equation}
\end{widetext}
If additional variational flexibility in the ansatz is desired, a
straightforward approach is to add additional layers of entanglers of the form
shown here, or to extend two-body entanglers to the next layer(s) of nearest
neighbors.

We now detail the specific choice of circuit for $\hat U_2$ used to cover
$SO(4)$ used in this work.  Note that there has been much interest in the
literature on the construction of optimal 2-body quantum circuits covering
$SU(4)$ or $SO(4)$ using various or arbitrary gate libraries - for an overview,
see \cite{Zhang:2003:027903,Shende:2004:062321,Vatan:2004:032315,Wei:2012:X}.
$SO(4)$ is the group of real, orthogonal matrices with determinant $+1$, and
covers all possible two-body entangler matrices needed in our VQE task. There
are infinitely many equivalent logical parametrizations of $SO(4)$ with 6 real
parameters, but some of these will prove easier to optimize and/or differentiate
than others in VQE applications.  For today, our selected two-body entangler
circuit is \cite{Wei:2012:X},
\begin{equation}
\label{eq:so4mark4}
\Qcircuit @R=1.0em @C=0.5em {
\lstick{|0_A\rangle}
 & \gate{R_{y} (\theta_{1})}
 & \ctrl{1}
 & \gate{R_{y} (\theta_{3})}
 & \ctrl{1}
 & \gate{R_{y} (\theta_{5})}
 & \qw \\
\lstick{|0_B\rangle}
 & \gate{R_{y} (\theta_{2})}
 & \targ
 & \gate{R_{y} (\theta_{4})}
 & \targ
 & \gate{R_{y} (\theta_{6})}
 & \qw \\
}
\end{equation}
This circuit was chosen because of its simple representation of parameters in
terms of $\hat R_y$ gates, which allows for direct implementation of our circuit
gradient recipe developed above.

Note that when joining individual $SO(4)$ entangler circuit elements, adjacent
$\hat R_y$ gates should be merged to minimize the number of VQE parameters and
avoid zero eigenvalues of the SA-VQE parameter Hessian (caused by redundancies
in the parameters). E.g., for a $N=4$ linear VQE circuit, the simply adjoined
circuit,
\begin{widetext}
\begin{equation}
\Qcircuit @R=1.0em @C=0.5em {
\lstick{|0_A\rangle}
 & \gate{R_{y} (\theta_{1})}
 & \ctrl{1}
 & \gate{R_{y} (\theta_{3})}
 & \ctrl{1}
 & \gate{R_{y} (\theta_{5})}
 & \qw & \qw & \qw & \qw & \qw
 & \qw \\
\lstick{|0_B\rangle}
 & \gate{R_{y} (\theta_{2})}
 & \targ
 & \gate{R_{y} (\theta_{4})}
 & \targ
 & \gate{R_{y} (\theta_{6})}
 & \gate{R_{y} (\theta_{1'})}
 & \ctrl{1}
 & \gate{R_{y} (\theta_{3'})}
 & \ctrl{1}
 & \gate{R_{y} (\theta_{5'})}
 & \qw \\
\lstick{|0_M\rangle}
 & \gate{R_{y} (\theta_{1''})}
 & \ctrl{1}
 & \gate{R_{y} (\theta_{3''})}
 & \ctrl{1}
 & \gate{R_{y} (\theta_{5''})}
 & \gate{R_{y} (\theta_{2'})}
 & \targ
 & \gate{R_{y} (\theta_{4'})}
 & \targ
 & \gate{R_{y} (\theta_{6'})}
 & \qw \\
\lstick{|0_N\rangle}
 & \gate{R_{y} (\theta_{2'})}
 & \targ
 & \gate{R_{y} (\theta_{4'})}
 & \targ
 & \gate{R_{y} (\theta_{6'})}
 & \qw & \qw & \qw & \qw & \qw
 & \qw \\
}
\end{equation}
should be merged to,
\begin{equation}
\label{eq:vqe_entangler_merged}
\Qcircuit @R=1.0em @C=0.5em {
\lstick{|0_A\rangle}
 & \gate{R_{y} (\theta_{1})}
 & \ctrl{1}
 & \gate{R_{y} (\theta_{3})}
 & \ctrl{1}
 & \gate{R_{y} (\theta_{5})}
 & \qw & \qw & \qw & \qw
 & \qw \\
\lstick{|0_B\rangle}
 & \gate{R_{y} (\theta_{2})}
 & \targ
 & \gate{R_{y} (\theta_{4})}
 & \targ
 & \gate{R_{y} (\theta_{7})}
 & \ctrl{1}
 & \gate{R_{y} (\theta_{3'})}
 & \ctrl{1}
 & \gate{R_{y} (\theta_{5'})}
 & \qw \\
\lstick{|0_M\rangle}
 & \gate{R_{y} (\theta_{1''})}
 & \ctrl{1}
 & \gate{R_{y} (\theta_{3''})}
 & \ctrl{1}
 & \gate{R_{y} (\theta_{8})}
 & \targ
 & \gate{R_{y} (\theta_{4'})}
 & \targ
 & \gate{R_{y} (\theta_{6'})}
 & \qw \\
\lstick{|0_N\rangle}
 & \gate{R_{y} (\theta_{2'})}
 & \targ
 & \gate{R_{y} (\theta_{4'})}
 & \targ
 & \gate{R_{y} (\theta_{6'})}
 & \qw & \qw & \qw & \qw 
 & \qw \\
}
\end{equation}
\end{widetext}
To verify the correctness of the analytical gradients, below we will also use
shortened versions of these entangler circuits designed to limit the variational
flexibility of the MC-VQE ansatz and  to increase the magnitude of the response
terms.

\subsubsection{Subspace Eigenstates}

The final MC-VQE eigenstates are determined by solving for the subspace
eigenvectors $V_{\Xi \Theta}$ and eigenvalues $E_{\Theta}$ by classically
solving the entangled, contracted subspace eigenproblem,
\begin{equation}
\sum_{\Xi'}
H_{\Xi \Xi'} V_{\Xi' \Theta}
=
V_{\Xi \Theta} E_{\Theta}
\ : \
\sum_{\Xi}
V_{\Xi \Theta}
V_{\Xi \Theta'}
=
\delta_{\Theta \Theta'}
\end{equation}
where the entangled, contracted subspace Hamiltonian is,
\begin{equation}
H_{\Xi \Xi'}
\equiv
\langle \Phi_{\Xi} | 
\hat U_{\mathrm{VQE}}^{\dagger} \hat H \hat U_{\mathrm{VQE}}
| \Phi_{\Xi'} \rangle
\end{equation}
This can be evaluated by partial tomography measurements on a quantum computer.
\[
2 H_{\Xi \neq \Xi'}
=
\left (
\langle \Phi_{\Xi} |
+
\langle \Phi_{\Xi'} |
\right )
\hat U_{\mathrm{VQE}}^{\dagger}
\hat H
\hat U_{\mathrm{VQE}}
\left (
| \Phi_{\Xi} \rangle
+
| \Phi_{\Xi'} \rangle
\right ) / 2
\]
\begin{equation}
\label{eq:Htrans}
-
\left (
\langle \Phi_{\Xi} |
-
\langle \Phi_{\Xi'} |
\right )
\hat U_{\mathrm{VQE}}^{\dagger}
\hat H
\hat U_{\mathrm{VQE}}
\left (
| \Phi_{\Xi} \rangle
-
| \Phi_{\Xi'} \rangle
\right ) / 2
\end{equation}
\[
=
\langle \chi_{\Xi \Xi'}^{+} | \hat U_{\mathrm{VQE}}^{\dagger} \hat H \hat U_{\mathrm{VQE}} |
\chi_{\Xi \Xi'}^{+} \rangle
-
\langle \chi_{\Xi \Xi'}^{-} | \hat U_{\mathrm{VQE}}^{\dagger} \hat H \hat U_{\mathrm{VQE}} |
\chi_{\Xi \Xi'}^{-} \rangle
\]
where the ``interfering'' contracted reference states are,
\begin{equation}
| \chi_{\Xi \Xi'}^{\pm} \rangle
\equiv
\sum_{I}
| I \rangle
\chi_{I\Xi \Xi'}^{\pm}
\equiv
\sum_{I}
| I \rangle
\frac{1}{\sqrt{2}}
\left [
C_{I\Xi}
\pm
C_{I\Xi'}
\right ]
\end{equation}
I.e., modified CIS states which can be prepared by the usual CIS state
preparation circuits with modified angles.

The procedure to determine the subspace Hamiltonian matrix elements and subspace
eigenstates is as follows: Define the ``interfering'' contracted reference state
coefficients as,
\begin{equation}
\chi_{I \Xi \Xi'}^{\pm}
\equiv
\frac{1}{\sqrt{2}}
\left [
C_{I\Xi}
\pm
C_{I\Xi'}
\right ]
\ \forall \ \Xi > \Xi'
\end{equation}
Define the corresponding CIS state preparation circuit angles,
\begin{equation}
\theta_{M} 
[
\chi_{I \Xi \Xi'}^{\pm}
]
\end{equation}
Compute the off-diagonal VQE-entangled contracted subspace Hamiltonian elements,
\begin{equation}
H_{\Xi \Xi'}
=
H_{\Xi' \Xi}
=
\left (
\varepsilon
[
 \theta_{g} ,
 \theta_M [  \chi_{I \Xi \Xi'}^{+}  ] 
]
\right .
\end{equation}
\[
\left .
-
\varepsilon
[
 \theta_{g} ,
 \theta_M [  \chi_{I \Xi \Xi'}^{-}  ] 
]
\right ) / 2
\ \forall \ \Xi > \Xi'
\]
Note that the diagonal subspace Hamiltonian matrix elements are available from
the state-averaged VQE optimization step above.  Finally, classically
diagonalize the subspace Hamiltonian to determine the subspace eigenvectors
$V_{\Xi \Theta}$ and MC-VQE eigenvalues $E_{\Xi}$,
\begin{equation}
\sum_{\Xi'}
H_{\Xi \Xi'} V_{\Xi' \Theta}
=
V_{\Xi \Theta} E_{\Theta}
\ : \
\sum_{\Xi}
V_{\Xi \Theta}
V_{\Xi \Theta'}
=
\delta_{\Theta \Theta'}
\end{equation}

\subsubsection{Generating State Coefficients}

For convenience, the MC-VQE states can be re-expressed in terms of ``generating
states'' $| \Gamma_{\Theta} \rangle$,
\begin{equation}
| \Psi_{\Theta} \rangle
\equiv
\hat U_{\mathrm{VQE}} | \Gamma_{\Theta} \rangle
\end{equation}
where,
\begin{equation}
| \Gamma_{\Theta} \rangle
\equiv
\sum_{I}
| I \rangle
\Gamma_{I\Theta}
\end{equation}
where,
\begin{equation}
\Gamma_{I\Theta}
\equiv
\sum_{\Xi}
C_{I\Xi}
V_{\Xi \Theta}
\end{equation}
Thus the generating states are rotations of the contracted reference states, and
can also be expressed in CIS form.

To utilize this expression, classically form the ``generating state coefficients,''
\begin{equation}
\Gamma_{I\Theta}
\equiv
\sum_{\Xi}
C_{I\Xi}
V_{\Xi \Theta}
\end{equation}
And then define the corresponding CIS state preparation circuit angles,
\begin{equation}
\theta_{M} 
[
 \Gamma_{I \Theta} 
]
\end{equation}
As a check, one should find,
\begin{equation}
E_{\Theta}
=
\varepsilon
[
 \theta_g ,
 \theta_{M} [  \Gamma_{I \Theta}  ] 
]
\end{equation}

\subsection{Quantum MC-VQE Gradient Stage}

\subsubsection{Formal Definition: MC-VQE Energy Gradient and Lagrangian}

\textit{Pauli-Basis Density Matrices:} For a given MC-VQE wavefunction $|
\Psi_{\Theta} \rangle$, the adiabatic state energy is,
\begin{equation}
E_{\Theta}
\equiv
\langle
\Psi_{\Theta} 
| 
\hat H
| 
\Psi_{\Theta} 
\rangle
\end{equation}
and our objective in this section is to form the ``relaxed'' one- and
two-particle density matrices in the Pauli basis, e.g.,
\begin{equation}
\gamma_{\mathcal{Z}_{A}}^{\Theta}
\equiv
\diff{E_{\Theta}}{\mathcal{Z}_{A}}
, \
\forall
\ A
\end{equation}
\begin{equation}
\Gamma_{\mathcal{ZZ}_{AA'}}^{\Theta}
\equiv
\diff{E_{\Theta}}{\mathcal{ZZ}_{AA'}}
, \
\forall
\ A, A' \ \in \ <A,A'>
\end{equation}
Once these are obtained, the total energy gradients can be obtained in the
classical portion of the algorithm by the chain rule,
\[
\total_{\zeta} E_{\Theta}
=
\mathcal{E}^{\zeta}
+ 
\sum_{A}
\mathcal{Z}_{A}^{\zeta}
\gamma_{\mathcal{Z}_{A}}^{\Theta}
+ 
\mathcal{X}_{A}^{\zeta}
\gamma_{\mathcal{X}_{A}}^{\Theta}
\]
\[
+
\frac{1}{2}
\sum_{<A, A'>}
\mathcal{XX}_{AA'}^{\zeta}
\Gamma_{\mathcal{XX}_{AA'}}^{\Theta}
+
\mathcal{XZ}_{AA'}^{\zeta}
\Gamma_{\mathcal{XZ}_{AA'}}^{\Theta}
\]
\begin{equation}
+
\mathcal{ZX}_{AA'}^{\zeta}
\Gamma_{\mathcal{ZX}_{AA'}}^{\Theta}
+
\mathcal{ZZ}_{AA'}^{\zeta}
\Gamma_{\mathcal{ZZ}_{AA'}}^{\Theta}
.
\end{equation}
Where, e.g., $\mathcal{Z}_{A}^{\zeta} \equiv \total_{\zeta} \mathcal{Z}_{A}$
is the gradient of the matrix element $\mathcal{Z}_{A}$ with respect to $\zeta$,
which can be determined in a final, classical step (itself requiring classical
response theory). Note that this gradient always appears contracted with a
density matrix quantity, which can often be used to improve efficiency.

The ``unrelaxed'' one- and two-particle density matrices are, e.g.,
\begin{equation}
\gamma_{\mathcal{Z}_{A}}^{\Theta,0}
\equiv
\pdiff{E_{\mathrm{\Theta}}}{\mathcal{Z}_{A}}
=
\langle 
\Psi_{\Theta}
|
\hat Z_{A}
|
\Psi_{\Theta}
\rangle
, \
\forall
\ A
\end{equation}
\[
\Gamma_{\mathcal{ZZ}_{AA'}}^{\Theta,0}
\equiv
\pdiff{E_{\mathrm{\Theta}}}{\mathcal{ZZ}_{AA'}}
=
\langle 
\Psi_{\Theta}
|
\hat Z_{A}
\otimes
\hat Z_{A'}
|
\Psi_{\Theta}
\rangle
,
\]
\begin{equation}
\forall
\ A, A' \ \in \ <A,A'>
\end{equation}
These carry the Hellmann-Feynman contributions to the gradient, but are missing
the MC-VQE wavefunction response contributions.

\textit{MC-VQE Quantum Lagrangian:} The corresponding Lagrangian has subspace
eigenstate, state-averaged VQE entangler, and contracted reference state
constraint terms,
\begin{equation}
\mathcal{L}_{\Theta}
=
E_{\Theta}
+
\underbrace{
\mathcal{L}_{\Theta}^{\mathrm{SE}}
+
\mathcal{L}_{\Theta}^{\mathrm{VQE}}
+
\mathcal{L}_{\Theta}^{\mathrm{CRS}}
}_{0}
\end{equation}
Here, the subspace eigenstate (SE) Lagrangian contribution is,
\[
\mathcal{L}_{\Theta}^{\mathrm{SE}}
\equiv
\sum_{\Xi' \Theta'}
\tilde \Xi_{\Xi' \Theta'}^{\Theta}
\underbrace{
\left [
\sum_{\Xi''}
H_{\Xi' \Xi''}
V_{\Xi'' \Theta'}
-
V_{\Xi' \Theta'}
E_{\Theta'}
\right ]
}_{
\partial {E_{\Theta'}} / \partial {V_{\Xi' \Theta'}}
\equiv
R_{\Xi' \Theta'}}
\]
\begin{equation}
:
\
E_{\Theta}
\equiv
\sum_{\Xi} 
\sum_{\Xi'}
V_{\Xi \Theta}
H_{\Xi \Xi'}
V_{\Xi' \Theta}
\end{equation}
The state-averaged VQE Lagrangian contribution is,
\begin{equation}
\mathcal{L}_{\Theta}^{\mathrm{VQE}}
\equiv
\sum_{g}
\tilde \theta_{g}^{\Theta}
\pdiff{\bar E}{\theta_{g}}
\end{equation}
The contracted reference state (CRS) Lagrangian contribution is,
\[
\mathcal{L}_{\Theta}^{\mathrm{CRS}}
\equiv
\sum_{I} 
\sum_{\Xi}
\tilde \Xi_{I\Xi}^{\Theta}
\underbrace{
\left [
\sum_{I'}
H_{I I'}
C_{I'\Xi}
-
C_{I\Xi}
E_{\Xi}^{\mathrm{CIS}}
\right ]
}_{
\partial {E_{\Xi}^{\mathrm{CIS}}} / \partial {C_{I \Xi}}
\equiv
R_{I \Xi}}
\]
\begin{equation}
:
\
E_{\Xi}^{\mathrm{CIS}}
\equiv
\sum_{I} 
\sum_{I'}
C_{I\Xi}
H_{I I'}
C_{J\Xi}
\end{equation}

Note the nesting of parametric dependencies in the various Lagrangian
contributions,
\begin{equation}
E_{\Theta} (
V_{\Xi \Theta},
\theta_{g},
C_{I \Xi}
)
\end{equation}
\begin{equation}
\mathcal{L}_{\Theta}^{\mathrm{SE}} (
V_{\Xi \Theta},
\theta_{g},
C_{I \Xi}
)
\end{equation}
\begin{equation}
\mathcal{L}_{\Theta}^{\mathrm{VQE}} (
\theta_{g},
C_{I \Xi}
)
\end{equation}
\begin{equation}
\mathcal{L}_{\Theta}^{\mathrm{CRS}} (
C_{I \Xi}
)
\end{equation}
This nesting is ubiquitous in Lagrangian approaches to analytical derivative
theory. When determining the base quantity $E_{\Theta}$, we start at the right
side of the Lagrangian and work left. When determining the gradient
$E_{\Theta}^\zeta$, we start at the left side of the Lagrangian and work right. 

\subsubsection{Subspace Eigenstate Response (Analytical)} 

The subspace eigenstate response equations have an analytical solution that
yields a trivial result for the special case of energy gradients. However, it is
still useful to explicitly demonstrate this, as nonzero responses arise from
this term for other properties such as non-adiabatic coupling vectors.

The subspace eigenstate response equations are,
\begin{equation}
\pdiff{\mathcal{L}_{\Theta}}{V_{\Xi' \Theta'}} = 0
\ \forall \ \Xi', \Theta'
\end{equation}
The ``driver'' term is,
\begin{equation}
E_{\Theta}
\equiv
\sum_{\Xi}
\sum_{\Xi'}
V_{\Xi \Theta}
H_{\Xi \Xi'}
V_{\Xi' \Theta}
\end{equation}
The gradient is,
\[
G_{\Xi' \Theta'}^{\Theta}
\equiv
\pdiff{E_{\Theta}}{V_{\Xi' \Theta'}}
=
2 \sum_{\Xi''}
H_{\Xi' \Xi''} 
V_{\Xi'' \Theta'}
\delta_{\Theta' \Theta}
\]
\begin{equation}
=
2
V_{\Xi' \Theta'}
E_{\Theta'}
\delta_{\Theta' \Theta}
\end{equation}
The Hessian is,
\begin{equation}
\mathcal{H}_{\Xi' \Theta', \Xi'' \Theta''}
\equiv
\pdiff{^2 E_{\Theta}}{
V_{\Xi' \Theta'}
\partial
V_{\Xi'' \Theta''}
}
\end{equation}
\[
=
\delta_{\Theta' \Theta''}
\left [
H_{\Xi' \Xi''}
-
E_{\Theta'}
\left (
\delta_{\Xi' \Xi''}
+
2
V_{\Xi' \Theta'}
V_{\Xi'' \Theta'}
\right )
\right ]
\]
Note that the Hessian is block diagonal in $\Theta$ and that the only nonzero
gradient term is in the specific state $\Theta$. Therefore, we will only need to
solve the subspace eigenstate response equations in the specific state $\Theta$.

The subspace eigenstate response equations (for state $\Theta$) are,
\begin{equation}
\sum_{\Xi'}
\mathcal{H}_{\Xi \Theta, \Xi' \Theta}
\tilde \Xi_{\Xi' \Theta}^{\Theta}
=
-
G_{\Xi \Theta}^{\Theta}
\end{equation}
By inspection (e.g., substitute into the above), the analytical solution is,
\begin{equation}
\tilde \Xi_{\Xi \Theta}^{\Theta}
=
V_{\Xi \Theta}
\end{equation}
Therefore, the subspace eigenstate response contribution is,
\[
\mathcal{L}_{\Theta}^{\mathrm{SE}}
\equiv
\sum_{\Xi}
\tilde \Xi_{\Xi \Theta}^{\Theta}
\left [
\sum_{\Xi'}
H_{\Xi \Xi'}
V_{\Xi' \Theta}
-
V_{\Xi \Theta}
E_{\Theta}
\right ]
\]
\begin{equation}
=
E_{\Theta}
-
E_{\Theta}
=
\boxed{
0
}
\end{equation}
Note that we are able to write the final zero in this context only because the
resultant Lagrangian quantity is absolutely zero in the sense that it and all
possible derivatives are zero. 

Therefore, we can entirely ignore contributions from
$\mathcal{L}_{\Theta}^{\mathrm{SE}}$ for the rest of the derivation of the
energy gradient.

\subsubsection{Unrelaxed Pauli Density Matrices}

Form the unrelaxed Pauli density matrices, e.g.,
\begin{equation}
\gamma_{\mathcal{Z}_{A}}^{\Theta,0}
\equiv
\pdiff{E_{\Theta}}{\mathcal{Z}_{A}}
=
\lambda_{\mathcal{Z}_{A}}
[ 
 \theta_g ,
 \theta_{M} [  \Gamma_{I \Theta}  ] 
]
\end{equation}
\begin{equation}
\Gamma_{\mathcal{ZZ}_{AB}}^{\Theta,0}
\equiv
\pdiff{E_{\Theta}}{\mathcal{ZZ}_{AB}}
=
\Lambda_{\mathcal{ZZ}_{AB}}
[ 
 \theta_g ,
 \theta_{M} [  \Gamma_{I \Theta}  ] 
]
\end{equation}

\subsubsection{CP-SA-VQE Response}

The coupled-perturbed state-averaged variational quantum eigensolver equations
(CP-SA-VQE) are,
\begin{equation}
\pdiff{\mathcal{L}_{\Theta}}{\theta_g} 
=
0
\ \forall \ g
\end{equation}
Both the right- and left-hand-sides of the CP-SA-VQE equations involve
quantities from quantum tomography measurements.
    
\subsubsection{CP-SA-VQE Response RHS}

Form the gradient of the energy with respect to VQE parameters,
\begin{equation}
G_{g}^{\Theta}
\equiv
\pdiff{E_{\Theta}}{\theta_{g}}
=
\varepsilon^{\theta_{g} + \pi / 4}
[
 \theta_{g} ,
 \theta_{M} [  \Gamma_{I \Theta}  ] 
]
\end{equation}
\[
-
\varepsilon^{\theta_{g} - \pi / 4}
[
 \theta_{g} ,
 \theta_{M} [  \Gamma_{I \Theta}  ] 
]
\]

\subsubsection{CP-SA-VQE Response LHS}

Form the SA-VQE Hessian, including the diagonal,
\[
\mathcal{H}_{gg}
\equiv
\pdiff{^2 \bar E}{\theta_g^2}
=
\frac{1}{N_{\Theta}}
\sum_{\Theta}
\varepsilon^{\theta_{g} + \pi / 2}
[
 \theta_{g} ,
 \theta_{M} [  C_{I \Theta}  ] 
]
\]
\[
+ 2
\varepsilon^{\theta_{g}}
[
 \theta_{g} ,
 \theta_{M} [  C_{I \Theta}  ] 
]
\]
\begin{equation}
-
\varepsilon^{\theta_{g} - \pi / 2}
[
 \theta_{g} ,
 \theta_{M} [  C_{I \Theta}  ] 
]
\end{equation}
and off-diagonal contributions,
\[
\mathcal{H}_{g\neq g'}
\equiv
\pdiff{^2 \bar E}{\theta_g \partial \theta_{g'}}
=
\frac{1}{N_{\Theta}}
\sum_{\Theta}
\varepsilon^{\theta_{g} + \pi / 4, \theta_{g'} + \pi /4}
[
 \theta_{g} ,
 \theta_{M} [  C_{I \Theta}  ] 
]
\]
\[
-
\varepsilon^{\theta_{g} + \pi / 4, \theta_{g'} - \pi /4}
[
 \theta_{g} ,
 \theta_{M} [  C_{I \Theta}  ] 
]
\]
\[
-
\varepsilon^{\theta_{g} - \pi / 4, \theta_{g'} + \pi /4}
[
 \theta_{g} ,
 \theta_{M} [  C_{I \Theta}  ] 
]
\]
\begin{equation}
\label{eq:vqe-hess}
+
\varepsilon^{\theta_{g} - \pi / 4, \theta_{g'} - \pi /4}
[
 \theta_{g} ,
 \theta_{M} [  C_{I \Theta}  ] 
]
\end{equation}

\subsubsection{CP-SA-VQE Response Equations}

Solve the CP-SA-VQE response equations,
\begin{equation}
\sum_{g'}
\mathcal{H}_{gg'}
\tilde \theta_{g'}^{\Theta}
=
-
G_{g}^{\Theta}
\end{equation}

\subsubsection{CP-SA-VQE Response Contribution}

The CP-SA-VQE response contribution to the Pauli density matrix is, e.g.,
\begin{equation}
\gamma_{\mathcal{Z}_{A}}^{\Theta, \mathrm{VQE}}
=
\sum_{g}
\tilde \theta_{g}
\pdiff{^2 \bar E}{\theta_{g} \partial \mathcal{Z}_{A}}
=
\frac{1}{N_{\Theta}}
\sum_{\Theta}
\sum_{g}
\tilde \theta_{g}
\end{equation}
\[
\left [
\lambda_{\mathcal{Z}_{A}}^{\theta_{g} + \pi / 4}
[
 \theta_{g} ,
 \theta_{M} [  C_{I \Theta}  ] 
]
\right .
\]
\[
\left .
-
\lambda_{\mathcal{Z}_{A}}^{\theta_{g} - \pi / 4}
[
 \theta_{g} ,
 \theta_{M} [  C_{I \Theta}  ] 
]
\right ]
\]

\subsubsection{CP-CRS Response}

The coupled-perturbed contracted reference state (CP-CRS) equations are,
\begin{equation}
\pdiff{\mathcal{L}_{\mathrm{\Theta}}}{C_{I \Xi}}
= 0
\ \forall \ I, \Xi
\end{equation}
The right-hand-side of the CP-CRS equations involves quantities from quantum
tomography measurements - the left-hand-side is a classical coupled-perturbed
configuration interaction singles (CP-CIS) Hessian.

\subsubsection{CP-CRS Response RHS}

The CP-CRS RHS is defined as,
\begin{equation}
G_{I \Xi}^{\Theta}
\equiv
\pdiff{E_{\Theta}}{C_{I \Xi}}
+
\pdiff{\mathcal{L}_{\Theta}^{\mathrm{VQE}}}{C_{I \Xi}}
\end{equation}

\subsubsection{CP-CRS Response RHS \#1}

The first contribution to the CP-CRS RHS involves the derivative of the state
energy with respect to the CIS CRS coefficients through the generator state
coefficients and angles,
\[
G_{I \Xi}^{\Theta}
\leftarrow
\sum_{M}
\pdiff{E_{\theta}}{\theta_{M} [ \Gamma_{I \Theta} ]}
\pdiff{\theta_{M} [ \Gamma_{I \Theta} ]}{\Gamma_{I \Theta}}
\pdiff{\Gamma_{I \Theta}}{C_{I \Xi}}
\]
\begin{equation}
=
\sum_{M}
\pdiff{\varepsilon[
 \theta_g ,
 \theta_{M} [ \Gamma_{I \Theta} ]
]}{
\theta_{M} [ \Gamma_{I \Theta} ]
}
\pdiff{\theta_{M} [ \Gamma_{I \Theta} ]}{\Gamma_{I \Theta}}
V_{\Xi \Theta}
\end{equation}
The gradients of the CIS circuit angles with respect to the corresponding CIS
coefficients, i.e., $\partial \theta_{M} / \partial \Gamma_{I \Theta}$ are
classically computed by the Jacobian formula in Equation \ref{eq:cis_jacobian}.

The gradients of the VQE-entangled CIS contracted reference states with respect
to CIS circuit angles, i.e., $\partial \varepsilon / \partial \theta_{M}$ can be
evaluated by analytical circuit gradients, but care must be taken to account for
the fact that pairs of $\hat R_y$ gates (plus a CZ gate or equivalent) are used
to implement the controlled $\hat F_y$ gate for CIS circuits as laid out in
Equations \ref{eq:cis_circuit} and \ref{eq:Fy_circuit}. The gradient for the
reference angle $\theta_{0}$ is the usual,
\begin{equation}
\pdiff{\varepsilon}{\theta_{0}}
=
\varepsilon^{\theta_{0} + \pi / 4}
-
\varepsilon^{\theta_{0} - \pi / 4}
\end{equation}
But the gradients the later angles such as $\theta_{AB}$ have additional contributions,
\[
\pdiff{\varepsilon}{\theta_{AB}}
=
-
(
\varepsilon^{\theta_{AB}^{\mathrm{L}} + \pi / 4}
-
\varepsilon^{\theta_{AB}^{\mathrm{L}} - \pi / 4}
) / 2
\]
\begin{equation}
+
(
\varepsilon^{\theta_{AB}^{\mathrm{R}} + \pi / 4}
-
\varepsilon^{\theta_{AB}^{\mathrm{R}} - \pi / 4}
) / 2
\end{equation}
Where $\theta_{AB}^{\mathrm{L}} \equiv - \theta_{AB} / 2$ is the argument to the
left-side $\hat R_y$ gate, and $\theta_{AB}^{\mathrm{R}} \equiv + \theta_{AB} /
2$ is the argument to the right-side $\hat R_y$ gate in Equation
\ref{eq:Fy_circuit}.

\subsubsection{CP-CRS Response RHS \#2}

The second contribution to the CP-CRS RHS involves the derivative of the SA-VQE
Lagrangian energy with respect to the CIS CRS coefficients,
\begin{equation}
\label{eq:cp-cis-rhs-2}
G_{I\Xi}^{\Theta}
\leftarrow
\sum_{g}
\sum_{M}
\tilde \theta_{g}^{\Theta}
\pdiff{^2 \bar E}{
\theta_{g} \partial \theta_{M} [C_{I \Xi}]}
\pdiff{\theta_{M} [C_{I \Xi}]}{C_{I \Xi}}
\end{equation}
\[
=
\frac{1}{N_{\Theta}}
\sum_{g}
\sum_{M}
\tilde \theta_{g}^{\Theta}
\pdiff{
\varepsilon^{\theta_{g} + \pi / 4}
[
 \theta_g ,
 \theta_{M} [C_{I \Xi}] 
]
}{\theta_{M} [C_{I \Xi}]}
\pdiff{\theta_{M} [C_{I \Xi}]}{C_{I \Xi}}
\]
\[
-
\frac{1}{N_{\Theta}}
\sum_{g}
\sum_{M}
\tilde \theta_{g}^{\Theta}
\pdiff{
\varepsilon^{\theta_{g} - \pi / 4}
[
 \theta_g ,
 \theta_{M} [C_{I \Xi}] 
]
}{\theta_{M} [C_{I \Xi}]}
\pdiff{\theta_{M} [C_{I \Xi}]}{C_{I \Xi}}
\]
This can be evaluated with the analytical CIS circuit gradients and CIS angle
Jacobian discussed in the previous section.

\subsubsection{CP-CRS Response LHS}

The CP-CRS Hessian is (block diagonal in CIS eigenstate $\Theta$),
\begin{equation}
\mathcal{H}_{I\Xi, I'\Xi'}
=
\delta_{\Xi \Xi'}
\left [
H_{II'}
-
E_{\Xi}^{\mathrm{CIS}}
\left (
\delta_{II'}
+
2 C_{I\Xi} C_{I' \Xi}
\right )
\right ]
\end{equation}

\subsubsection{CP-CRS Response Equations}

Solve the CP-CRS response equations (block diagonal in CIS eigenstate $\Xi$),
\begin{equation}
\mathcal{H}_{I\Xi, I'\Xi'}
\tilde \Xi_{I' \Xi'}^{\Theta}
=
- G_{I \Xi}^{\Theta}
\end{equation}

\subsubsection{CP-CRS Response Contribution}

The CRS response contribution to the density matrix is, e.g.,
\[
\gamma_{\mathcal{Z}_{A}}^{\Theta,\mathrm{CRS}}
\equiv
\pdiff{}{\mathcal{Z}_{A}}
\sum_{I\Xi}
\tilde \Xi_{I\Xi}^{\Theta}
R_{I\Xi}
=
\sum_{I\Xi}
\tilde \Xi_{I\Xi}^{\Theta}
\pdiff{}{\mathcal{Z}_{A}}
R_{I\Xi}
\]
\[
=
\sum_{I\Xi}
\sum_{I' I''}
\tilde \Xi_{I\Xi}^{\Theta}
\pdiff{R_{I\Xi}}{H_{I'I''}}
\pdiff{H_{I'I''}}{\mathcal{Z}_{A}}
\]
\begin{equation}
=
\sum_{I'I''}
D_{I'I''}^{\Theta}
\pdiff{H_{I'I''}}{\mathcal{Z}_{A}}
\end{equation}
where,
\begin{equation}
D_{I'I''}^{\Theta}
=
\sum_{\Xi}
\tilde \Xi_{I' \Xi}^{\Theta}
C_{I'' \Xi}
-
\sum_{\Xi}
\left [
\sum_{I}
\tilde \Xi_{I \Xi}^{\Theta}
C_{I \Xi}
\right ]
C_{I' \Xi}
C_{I'' \Xi}
\end{equation}
The final partials of the CIS Hamiltonian with respect to the Pauli-basis
potential matrix elements, e.g., $\partial_{\mathcal{Z}_{A}} H_{I'I''}$, are easily
formed by differentiating Equations \ref{eq:cis_H_00} to \ref{eq:cis_H_CD},
which are linear in the Pauli basis potential matrix elements.

\subsubsection{Relaxed Pauli Density Matrices}

The target relaxed Pauli density matrices are simply given as the sum of the
unrelaxed density matrices plus the individual response contributions, e.g.,
\begin{equation}
\label{eq:pauli_dm_relaxed}
\gamma_{\mathcal{Z}_{A}}^{\Theta}
\equiv
\diff{E_{\Theta}}{\mathcal{Z}_{A}}
=
\pdiff{\mathcal{L}_{\Theta}}{\mathcal{Z}_{A}}
=
\gamma_{\mathcal{Z}_{A}}^{\Theta,0}
+
\gamma_{\mathcal{Z}_{A}}^{\Theta,\mathrm{VQE}}
+
\gamma_{\mathcal{Z}_{A}}^{\Theta,\mathrm{CRS}}
\end{equation}

\subsection{Classical AIEM Gradient Stage}

\subsubsection{Pauli Basis AIEM Gradient}

At present, we have the Lagrangian representation,
\[
E_{\Theta}
=
\mathcal{L}_{\Theta}
=
\mathcal{E}
+ 
\sum_{A}
\mathcal{Z}_{A}
\gamma_{\mathcal{Z}_{A}}^{\Theta}
+ 
\mathcal{X}_{A}
\gamma_{\mathcal{X}_{A}}^{\Theta}
\]
\[
+
\frac{1}{2}
\sum_{<A, A'>}
\mathcal{XX}_{AA'}
\Gamma_{\mathcal{XX}_{AA'}}^{\Theta}
+
\mathcal{XZ}_{AA'}
\Gamma_{\mathcal{XZ}_{AA'}}^{\Theta}
\]
\begin{equation}
+
\mathcal{ZX}_{AA'}
\Gamma_{\mathcal{ZX}_{AA'}}^{\Theta}
+
\mathcal{ZZ}_{AA'}
\Gamma_{\mathcal{ZZ}_{AA'}}^{\Theta}
\end{equation}
But, due to the solution of the response equations above, the Lagrangian is
fully relaxed in terms of MC-VQE parameters. Therefore,
\[
\total_{\zeta} E_{\Theta}
=
\partial_{\zeta} \mathcal{L}_{\Theta}
=
\mathcal{E}^{\zeta}
+ 
\sum_{A}
\mathcal{Z}_{A}^{\zeta}
\gamma_{\mathcal{Z}_{A}}^{\Theta}
+ 
\mathcal{X}_{A}^{\zeta}
\gamma_{\mathcal{X}_{A}}^{\Theta}
\]
\[
+
\frac{1}{2}
\sum_{<A, A'>}
\mathcal{XX}_{AA'}^{\zeta}
\Gamma_{\mathcal{XX}_{AA'}}^{\Theta}
+
\mathcal{XZ}_{AA'}^{\zeta}
\Gamma_{\mathcal{XZ}_{AA'}}^{\Theta}
\]
\begin{equation}
+
\mathcal{ZX}_{AA'}^{\zeta}
\Gamma_{\mathcal{ZX}_{AA'}}^{\Theta}
+
\mathcal{ZZ}_{AA'}^{\zeta}
\Gamma_{\mathcal{ZZ}_{AA'}}^{\Theta}
\end{equation}
Where, e.g., $\mathcal{Z}_{A}^{\zeta} \equiv \total_{\zeta} \mathcal{Z}_{A}$
is the gradient of the matrix element $\mathcal{Z}_{A}$ with respect to $\zeta$.
What remains is to evaluate the derivatives of the classical matrix elements
(including any classical response equations) and efficiently assemble the
finished analytical derivative $\total_{\zeta} E_{\Theta}$.

\subsubsection{Monomer Basis AIEM Gradient}

At this point it is advantageous to perform a linear transformation to a
monomer-basis matrix element representation of the gradient,
\[
\total_{\zeta} E_{\Theta}
=
\sum_{A}
\epsilon_{\mathrm{H}}^{A,\zeta}
\gamma_{\mathrm{H}}^{A,\Theta}
+
\epsilon_{\mathrm{P}}^{A,\zeta}
\gamma_{\mathrm{P}}^{A,\Theta}
+
\underbrace{
\epsilon_{\mathrm{T}}^{A,\zeta}
}_{0}
\gamma_{\mathrm{T}}^{A,\Theta}
\]
\[
+
\frac{1}{2}
\sum_{<A,A'>}
v_{\mathrm{HH}}^{AA',\zeta}
\Gamma_{\mathrm{HH}}^{AA',\Theta}
+
v_{\mathrm{HT}}^{AA',\zeta}
\Gamma_{\mathrm{HT}}^{AA',\Theta}
+
v_{\mathrm{HP}}^{AA',\zeta}
\Gamma_{\mathrm{HP}}^{AA',\Theta}
\]
\[
+
v_{\mathrm{TH}}^{AA',\zeta}
\Gamma_{\mathrm{TH}}^{AA',\Theta}
+
v_{\mathrm{TT}}^{AA',\zeta}
\Gamma_{\mathrm{TT}}^{AA',\Theta}
+
v_{\mathrm{TP}}^{AA',\zeta}
\Gamma_{\mathrm{TP}}^{AA',\Theta}
\]
\begin{equation}
+
v_{\mathrm{PH}}^{AA',\zeta}
\Gamma_{\mathrm{PH}}^{AA',\Theta}
+
v_{\mathrm{PT}}^{AA',\zeta}
\Gamma_{\mathrm{PT}}^{AA',\Theta}
+
v_{\mathrm{PP}}^{AA',\zeta}
\Gamma_{\mathrm{PP}}^{AA',\Theta}
\end{equation}
Here, the monomer-basis density matrices are defined as, e.g.,
\[
\gamma_{\mathrm{H}}^{A'',\Theta}
\equiv
\pdiff{}{\epsilon_{\mathrm{H}}^{A''}}
\mathcal{L}_{\Theta}
=
\pdiff{}{\epsilon_{\mathrm{H}}^{A''}}
\left [
\mathcal{E}
+
\sum_{A}
\mathcal{Z}_{A}
\gamma_{\mathcal{Z}_{A}}^{\Theta}
+ 
\ldots
\right ]
\]
\begin{equation}
\label{eq:monomer_dm_e}
=
\underbrace{
\left [
\pdiff{}{\epsilon_{\mathrm{H}}^{A''}}
\mathcal{E}
\right ]
}_{1/2}
+
\sum_{A}
\underbrace{
\left [
\pdiff{}{\epsilon_{\mathrm{H}}^{A''}}
\mathcal{Z}_{A}
\right ]
}_{\delta_{AA''} / 2}
\gamma_{\mathcal{Z}_{A}}^{\Theta}
+ 
\ldots
\end{equation}
The transformation coefficients indicated by underbraces are easily determined
by differentiating Equations \ref{eq:aiem_pauli_E} to \ref{eq:aiem_pauli_ZZ},
which are linear in the monomer-basis potential matrix elements.

\subsubsection{Dimer Interaction Matrix Element Gradients}

At this point, it is advantageous to contract through partial derivatives of
the dimer interaction matrix elements to obtain a monomer-property-only
representation of the analytical derivative.
\[
\total_{\zeta} E_{\Theta}
=
\sum_{A}
\epsilon_{\mathrm{H}}^{A,\zeta}
\gamma_{\mathrm{H}}^{A,\Theta}
+
\epsilon_{\mathrm{P}}^{A,\zeta}
\gamma_{\mathrm{P}}^{A,\Theta}
+
\underbrace{
\epsilon_{\mathrm{T}}^{A,\zeta}
}_{0}
\gamma_{\mathrm{T}}^{A,\Theta}
\]
\begin{equation}
\label{eq:aiem_grad_mon_prop}
+
\vec \mu_{\mathrm{H}}^{A,\zeta}
\vec \eta_{\mathrm{H}}^{A,\Theta}
+
\vec \mu_{\mathrm{P}}^{A,\zeta}
\vec \eta_{\mathrm{P}}^{A,\Theta}
+
\vec \mu_{\mathrm{T}}^{A,\zeta}
\vec \eta_{\mathrm{T}}^{A,\Theta}
+
\vec r_{0}^{A,\zeta}
\vec \xi^{\mathrm{A,\Theta}}
\end{equation}

The monomer property density matrices are defined as, e.g.,
\[
\vec \eta_{\mathrm{H}}^{A'', \Theta}
\equiv
\pdiff{}{\vec \mu_{\mathrm{H}}^{A''}}
\mathcal{L}_{\Theta}
=
\pdiff{}{\vec \mu_{\mathrm{H}}^{A''}}
\frac{1}{2}
\sum_{<A,A'>}
v_{\mathrm{HH}}^{AA'}
\Gamma_{\mathrm{HH}}^{AA',\Theta}
\ldots
\]
\begin{equation}
\label{eq:monomer_dm_mu}
=
\frac{1}{2}
\sum_{<A,A'>}
\left [
\pdiff{}{\vec \mu_{\mathrm{H}}^{A''}}
v_{\mathrm{HH}}^{AA'}
\right ]
\Gamma_{\mathrm{HH}}^{AA',\Theta}
\ldots
\end{equation}
and,
\[
\vec \xi^{A'', \Theta}
\equiv
\pdiff{}{\vec r_{0}^{A''}}
\mathcal{L}_{\Theta}
=
\pdiff{}{\vec r_{0}^{A''}}
\frac{1}{2}
\sum_{<A,A'>}
v_{\mathrm{HH}}^{AA'}
\Gamma_{\mathrm{HH}}^{AA',\Theta}
\ldots
\]
\begin{equation}
\label{eq:monomer_dm_r0}
=
\frac{1}{2}
\sum_{<A,A'>}
\left [
\pdiff{}{\vec r_{0}^{A''}}
v_{\mathrm{HH}}^{AA'}
\right ]
\Gamma_{\mathrm{HH}}^{AA',\Theta}
\ldots
\end{equation}

Some needed partials of Equation \ref{eq:v_dipole_dipole} are,
\[
\pdiff{}{\vec \mu_{A''}}
v^{AA'}
=
\left [
\frac{
\vec \mu_{A'}
}{r_{AA'}^3}
-
3
\frac{
(
\vec \mu_{A'}
\cdot
\vec r_{AA'}
)
\vec r_{AA'}
}{r_{AA'}^5}
\right ]
\delta_{AA''}
\]
\begin{equation}
+
\left [
\frac{
\vec \mu_{A}
}{r_{AA'}^3}
-
3
\frac{
(
\vec \mu_{A}
\cdot
\vec r_{AA'}
)
\vec r_{AA'}
}{r_{AA'}^5}
\right ]
\delta_{A'A''}
\end{equation}
and,
\[
\pdiff{}{\vec r_{A''}^{0}}
v^{AA'}
=
\left [
-3
\frac{
(
\vec \mu_{A}
\cdot
\vec \mu_{A'}
)
\vec r_{AA'}
}{r_{AA'}^5}
\right .
\]
\[
\left .
+15
\frac{
(\vec \mu_{A} \cdot \vec r_{AA'})
(\vec \mu_{A'} \cdot \vec r_{AA'})
\vec r_{AA'}
}{r_{AA'}^7}
\right .
\]
\begin{equation}
\label{eq:v_grad_r0}
\left .
-3
\frac{
(\vec \mu_{A'} \cdot \vec r_{AA'})
\vec \mu_{A}
}{r_{AA'}^5}
-3
\frac{
(\vec \mu_{A} \cdot \vec r_{AA'})
\vec \mu_{A'}
}{r_{AA'}^5}
\right ]
\left [
\delta_{A'A''}
-
\delta_{AA''}
\right ]
\end{equation}

\subsubsection{Monomer Property Gradients}

Finally, the contraction of the monomer property density matrices with the
monomer property nuclear gradients can be performed as indicated in Equation
\ref{eq:aiem_grad_mon_prop}. For today's exercise, this is a simple set of
multiply-add operations with pre-supplied monomer property gradients (obtained
with classical Lagrangian theory in methods like TD-DFT, including classical
response equations like CP-KS). However, it is also possible to defer
computation of these derivatives until the last possible moment, at which point
the monomer property density matrices are known, and can be contracted with the
monomer property gradients on the fly. This can reduce the number of classical
response equations that must be solved, and can help sieve the classical
gradient equations if some of the monomer property density matrices are sparse.

\section{Computational Details}

The MC-VQE+AIEM energy and analytical gradient expressions of Equations
\ref{eq:v_dipole_dipole} through \ref{eq:v_grad_r0} were implemented in our
in-house quantum circuit simulator \textsc{Quasar}. For the purposes of this
manuscript, we simulate ideal infinite sampling of Pauli density matrices by
contraction of the relevant statevector amplitudes.  Ground state properties are
computed with Kohn-Sham density functional theory (KS-DFT), excited state
properties are computed with time-dependent DFT in the Tamm-Dancoff
approximation (TDA-TD-DFT).\cite{Runge:1984:997} The excited state and transition dipole moments are
computed in the unrelaxed expectation value formalism. Dipole moments are
computed at the mass centers of the monomers. Analytical gradients of the
ground- and excited-state energies and the ground-state, excited-state and
transition dipole moments are computed within a classical Lagrangian approach
involving coupled-perturbed configuration interaction (CP-CI) and
coupled-perturbed Kohn Sham (CP-KS) equations in the relevant places. These
derivative quantities are computed up front, and are stored for later
contraction with the relaxed density matrices emanating from the quantum MC-VQE
portion of the algorithm - we do not explore the full Lagrangian approach with
delayed computation of classical response in the present work. All classical
electronic structure computations are performed in double precision (unless
otherwise indicated) within the \textsc{TeraChem} GPU-accelerated electronic
structure package,\cite{Ufimtzsev:2009:2619,Luehr:2011:949,Isborn:2012:5092} and all iterative portions of the computations are tightly
converged to facilitate comparison of numerical derivative properties.

\section{Results and Discussion}

\subsection{Validation: BChl-a Dimer}

\label{sec:validation}

\begin{figure}[h!]
\begin{center}
\includegraphics[width=3.1in]{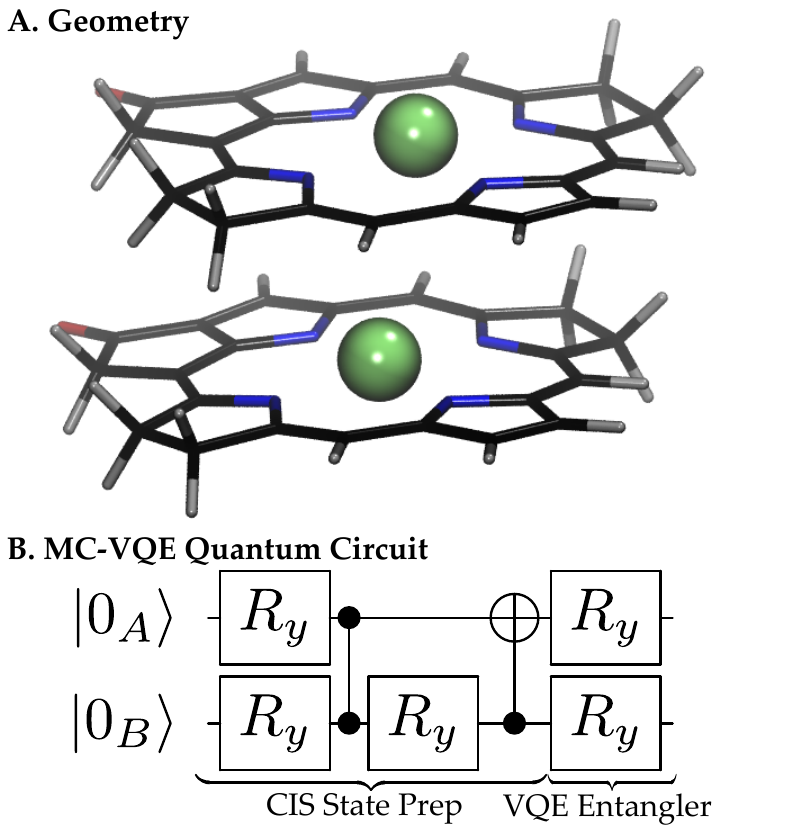}
\caption{Schematic of (A) BChl-a dimer system and (B) corresponding truncated
MC-VQE quantum circuit selected for analytical gradient validation and
excited-state dynamics demonstration in Sections \ref{sec:validation} and
\ref{sec:dynamics}. The VQE entangler circuit element is deliberately truncated
to limit the flexibility of the ansatz, yielding large wavefunction response
contributions.}
\label{fig:dimer-schem}
\end{center}
\end{figure}

As implied by the verbosity of the derivation above, analytical derivative
theory requires extensive demonstration of validity to eliminate myriad
potential sources of mathematical or programmatical errors. To proceed, we
consider the case of a small dimer of simplified BChl-a chromophores depicted in
Figure \ref{fig:dimer-schem}. We shorten the VQE entangler circuit to only a
pair of $R_y$ gates and we consider only two contracted reference states out of
a total of three CIS states. This deliberately limits the flexibility and
accuracy of the MC-VQE ansatz, and thereby ensures that the contributions from
both CP-SA-VQE response and CP-CIS response will be large, providing a stress
test that the non-Hellmann-Feynman contributions have been correctly derived and
implemented.  We then compare the analytical derivatives to second-order
symmetric finite difference derivatives in three successive bases: the relaxed
Pauli density matrices \ref{eq:pauli_dm_relaxed} (the total derivatives of the
state energies with respect to perturbations in the Pauli Hamiltonian matrix
elements), the monomer property density matrices of Equations
\ref{eq:monomer_dm_e}, \ref{eq:monomer_dm_mu}, and \ref{eq:monomer_dm_r0} (the
total derivatives of the state energies with respect to perturbations in the
classical monomer properties), and the full nuclear gradient of Equation
\ref{eq:aiem_grad_mon_prop} (the total derivatives of the state energy with
respect to perturbations in the nuclear positions). This pipeline of comparisons
allows us to separately verify the gradients of the quantum portions of the
MC-VQE method, the gradients of the AIEM-to-Pauli Hamiltonian transformation,
and the classical electronic structure gradients.

\subsubsection{Pauli Density Matrices}

Table \ref{tab:pauli_dm} compares the deviations of Pauli-basis density matrices
computed with different methods and gradient approaches for the ground state and
first excited states of BChl-a dimer within the AIEM based on
$\omega$PBE($\omega=0.3$)/6-31G*-D3.\cite{Tawada:2004:8425,Vydrov:2006:234109}.
The principal finding is that MC-VQE including all response terms [denoted as
VQE(Y,Y) in the table] agrees to $\sim 10^{-10}$ with second-order symmetric
finite difference with a stepsize of $10^{-7}$ a.u. This is the same level of
agreement as obtained between the analytical gradient methodology and finite
difference within FCI and CIS, and is several orders of magnitude better
agreement than is obtained if either CP-SA-VQE or CP-CIS response is neglected
in MC-VQE. In fact, inspection of the top three and bottom three rows of the
table indicates that the errors induced by neglecting response terms in MC-VQE
gradients are of essentially the same order as the intrinsic deviations between
different methods. 

\subsubsection{Monomer Property Density Matrices}

Table \ref{tab:monomer_dm} shows the same analysis as Table \ref{tab:pauli_dm},
but now for monomer-property density matrices. The findings are identical - the
response-including MC-VQE gradients agree with finite difference to the same
order as FCI or CIS, and the errors induced by neglecting response terms in
MC-VQE gradients are of essentially the same order as the intrinsic deviations
between different methods. 

\subsubsection{Full Nuclear Gradients}

Table \ref{tab:nuclear_dm} shows the same analysis as Table
\ref{tab:monomer_dm}, but now for the complete nuclear gradients of selected
atoms. The very nature of this study highlights the imperative to have efficient
analytical gradient methodology where only a minimal number of response
equations are solved: obtaining the finite difference gradient for the full
system would have necessitated the computation of the classical electronic
structure quantities on a 528-point stencil, which is \emph{more computational
effort} than will be required in the excited-state dynamics study of the
following section. This effort was deemed to be overtly laborious, and so
finite-difference spot checks of the gradients were performed for the five
selected atoms indicated in the table. The conclusion is the same as in the
previous two sections - response-including MC-VQE gradients agree with finite
difference to the same order as FCI or CIS, and the errors induced by neglecting
response terms in MC-VQE gradients are of essentially the same order as the
intrinsic deviations between different methods. In all cases, the analytical
gradient methods agree with finite difference to at least one order of magnitude
better than if any response terms are neglected. Moreover, most of the remaining
deviation between the analytical gradient methodology and the finite difference
methodology likely originates from the error in the finite difference stencil.
As evidence for this, full analytical nuclear gradients for FCI and CIS were
checked between this code and a separate code written by a different author and
using a different set of working equations, and were found to agree to $\sim
10^{-15}$. This, together with the fact that the response including monomer
property density matrices of Table \ref{tab:monomer_dm} achieved a
more-substantial $\sim 10^{6}\times$
improvement in agreement with finite difference and the fact that the
contraction with the classical electronic structure property gradients between
Table \ref{tab:monomer_dm} and Table \ref{tab:nuclear_dm} is identical between
FCI, CIS, and MC-VQE, provides a strong indicator that the full MC-VQE gradients
are true analytical derivatives of the MC-VQE energy.

\begin{figure}[h!]
\begin{center}
\includegraphics[width=3.1in]{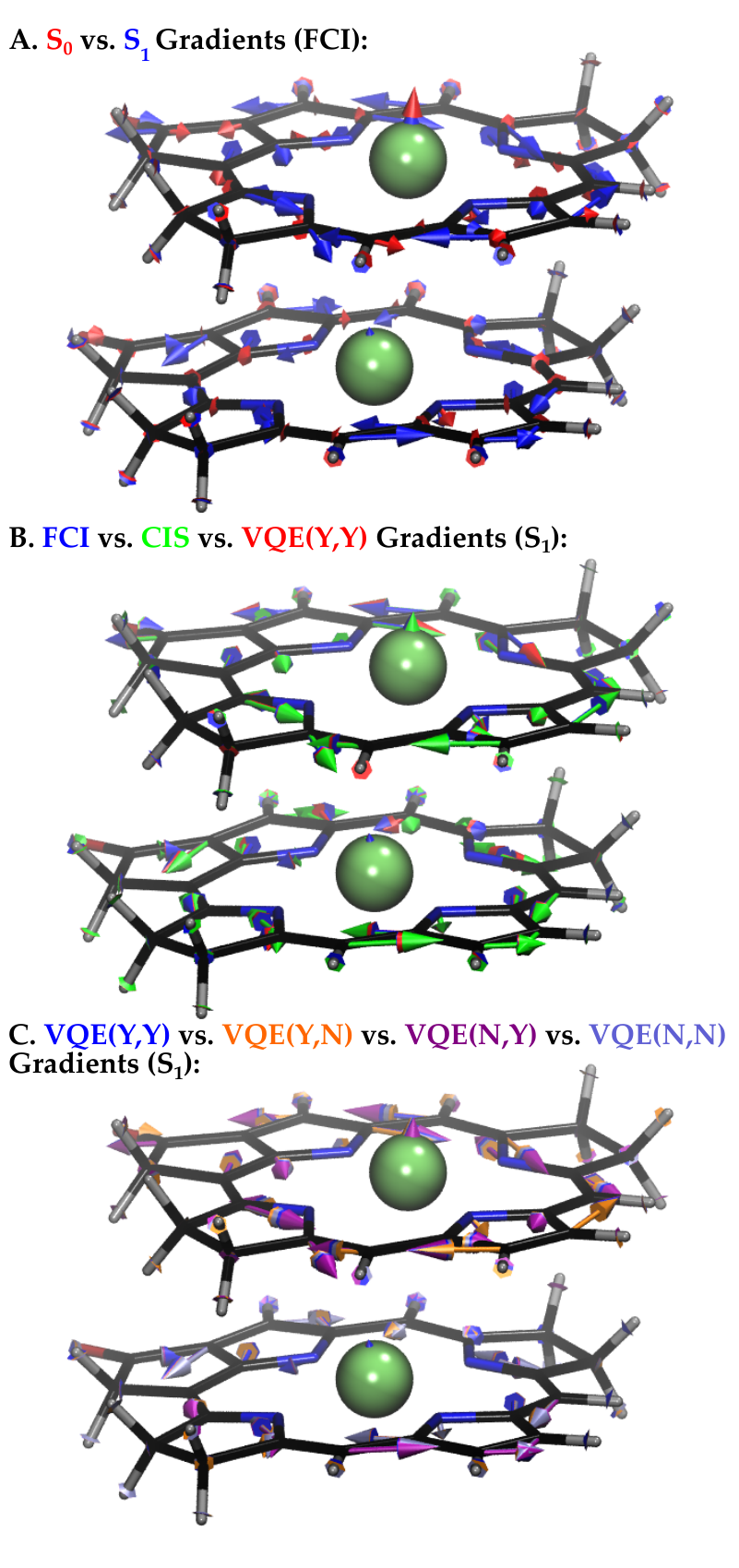}
\caption{
Visual comparison of nuclear gradients computed for BChl-a dimer within the AIEM
based on $\omega$PBE($\omega=0.3$)/6-31G*-D3. (A) Ground-state [red] vs.
first-excited-state [blue] gradients computed with FCI. (B) FCI [blue] vs. CIS
[green] vs. VQE [red] gradients computed on $S_1$. (C) VQE gradients computed
with different inclusion of response terms on $S_1$. The notation VQE(N,Y)
indicates that VQE response (the first argument) was not used while CIS response
(the second argument) was used.  For MC-VQE, two states were computed and the
VQE entangler circuit of Figure \ref{fig:dimer-schem} was used.}
\label{fig:grad-viz}
\end{center}
\end{figure}

Figure \ref{fig:grad-viz} shows a visual comparison of nuclear gradients
computed with various levels of theory and inclusions of response terms. The key
finding of this study is that there are visual deviations between MC-VQE
gradients using different levels of completeness of response terms (bottom
panel) and that these deviations are of the same scale as the intrinsic
deviations between different methods (top panel).  

\subsection{Demonstration: Excited State Dynamics}

\label{sec:dynamics}

\begin{figure}[h!]
\begin{center}
\includegraphics[width=3.1in]{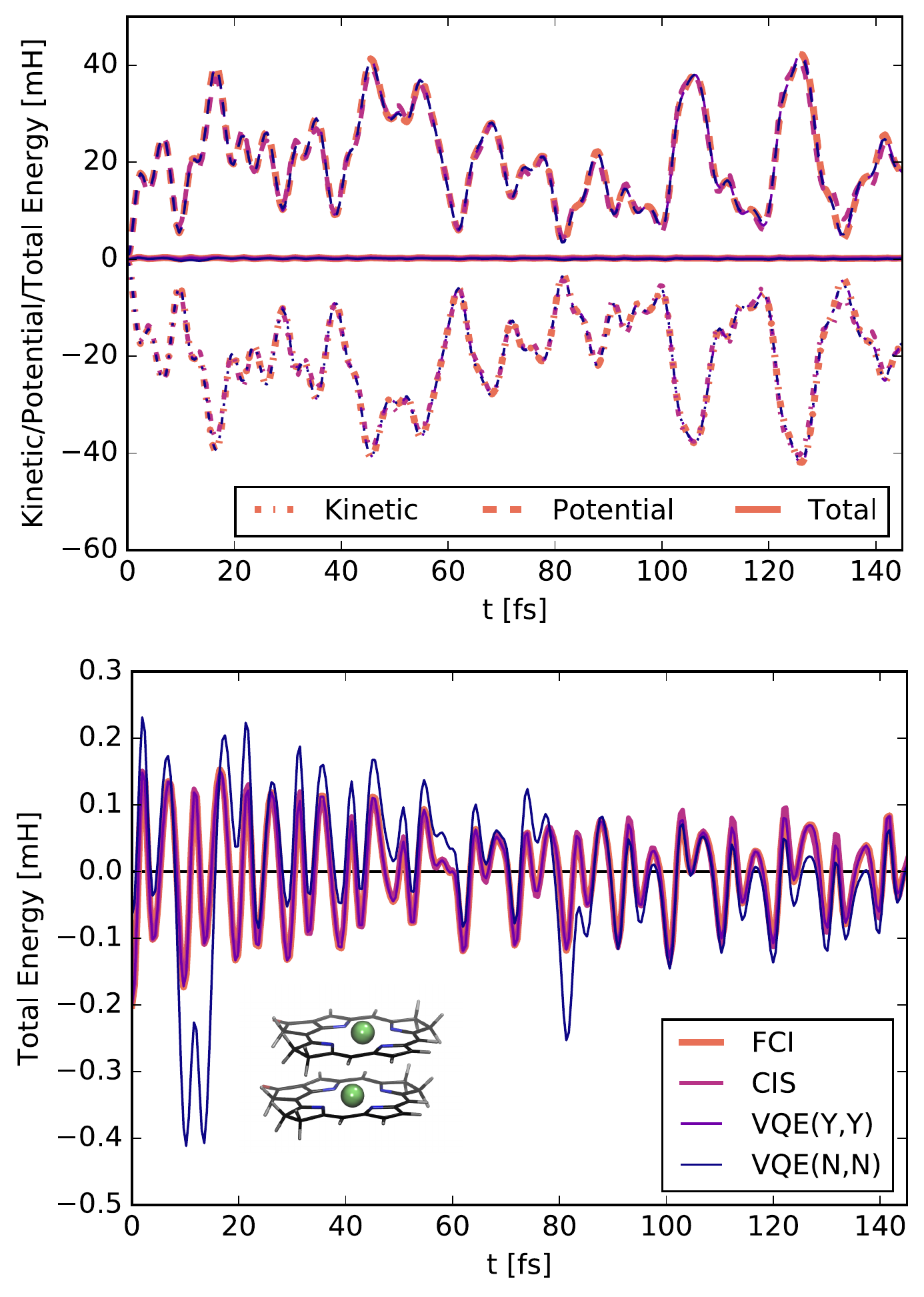}
\caption{Energy characteristics of $S_1$ adiabatic excited-state dynamics of
BChl-a dimer within the AIEM based on $\omega$PBE($\omega=0.3$)-D3/STO-3G,
computed using velocity Verlet with a timestep of 20 au ($\sim 0.5$ fs) and
analytical nuclear gradients. Top panel: Kinetic, potential, and total energy
profiles, all zeroed to the values at the initial frame. Bottom panel: magnified
view of the total energy profile from the top panel, zeroed to the mean total
energy. The notation VQE(N,Y) indicates that VQE response (the first argument)
was not used while CIS response (the second argument) was used.  For MC-VQE, two
states were computed and the VQE entangler circuit of Figure
\ref{fig:dimer-schem} was used.}
\label{fig:dimer-energy}
\end{center}
\end{figure}

To demonstrate the potential utility of our hybrid quantum/classical methodology,
we have performed an adiabatic excited state dynamics simulation of the system
shown in Figure \ref{fig:dimer-schem}. The methodology for this study is the
same as previously described, except that the basis is reduced to STO-3G and
mixed precision computations are used to facilitate computational expediency of
the monomer classical electronic structure computations. The dynamics are
initiated on $S_1$ with a randomized initial momentum vector drawn from a
Maxwellian distribution at $T=300$ K (the same initial conditions are used for
all methods). The dynamics are propagated using velocity Verlet with a timestep
of $20$ a.u. ($\sim 0.5$ fs) for 300 timesteps ($\sim 145$ fs). The energy
profiles of the dynamics are depicted in Figure \ref{fig:dimer-energy}.
Inspection of the top panel shows a rather ordinary adiabatic dynamics energy
profile with reasonable energy conservation profile considering the short
overall timescale of the dynamics. The bottom panel shows a magnified view of
the total energy profile, in this case zeroed to the mean total energy. A highly
compelling finding emerges: the energy conservation profiles of FCI, CIS, and
MC-VQE with full inclusion of response are essentially coincident to the order
of the linewidths (there are some minor deviations apparent if one zooms in),
but MC-VQE without response contributions immediately diverges from the other
curves and additionally exhibits larger ``excursions'' at $\sim 10$ fs and $\sim
75$ fs. Even though FCI, CIS, and MC-VQE all have intrinsic differences, it
seems that the internal self-consistency of their analytical gradients works to
promote a highly coincident energy conservation profile. By contrast, MC-VQE
without response contributions has a non-self-consistent gradient that manifests
as a loosely random perturbing force, leading to an immediate divergence of the
corresponding energy conservation profile. The energy excursions experienced by
MC-VQE without response are also characteristic of a non-self-consistent
gradient, and will likely lead to more-substantive energy conservation
problems at longer timescales. While more-flexible VQE entangler circuits will
doubtless lead to higher-accuracy MC-VQE solutions, and, therefore, smaller
response contributions, this study highlights the importance of fully
self-consistent analytical derivative methodology with complete inclusion of
wavefunction response. 

\subsection{Larger Systems and Entangler Circuits}

It might be reasonable to expect that one could forgo computation of the
response terms if a sufficiently flexible VQE entangler circuit were utilized -
i.e., in a hypothetical ``full MC-VQE'' ansatz, the Hellmann-Feynman formula
would be exact because the full MC-VQE states would be equivalent to the FCI
states, and one could eschew the response computation without loss. Here, we
provide a short case study that indicates that this is not generally the case
with practically-sized VQE entangler circuits. We consider the case of a linear
BChl-a hexamer ($N=6$), and use both a truncated VQE entangler circuit
consisting of a single $R_y$ gate at the end of each qubit wire (analogous to
the truncated entangler used for he BChl-a dimer test case above) and a
``standard'' VQE entangler circuit with $SO(4)$ entanglers between each nearest
neighbor as in Equation \ref{eq:vqe_entangler_merged}. The results are
summarized in Table \ref{tab:large}. The top portion of the table shows that the
use of the more-powerful standard VQE entangler circuit reduces the error of the
observables by $\sim 3-5\times$, concordant with a $4.4\times$ reduction in the
error in the full MC-VQE gradient in the fourth line of the bottom portion of
the table. This improvement moves the MC-VQE results from being negligibly
better than CIS with the truncated VQE entangler to substantially better than
CIS with the standard VQE entangler. However, the magnitudes of the response
terms in MC-VQE do not substantially diminish, as indicated in the first three
lines of the bottom portion of the table. In fact, for the standard entangler,
the response terms are of the same size or even larger than the discrepancy
between MC-VQE and FCI. Clearly, this is just one case study, and a
more-thorough analysis is warranted in future work, but these initial
results do seem to indicate that the response terms are non-negligible even
in larger systems with practically-sized (non-truncated) VQE entangler circuits.

\subsection{Prospects for Iterative Solution of Quantum Response Equations}

As mentioned in the introduction, one of the key findings of Lagrangian-based
analytical nuclear theory in classical electronic structure is that the nuclear
gradient (or other first derivative properties) can generally be formed in
effort that is strictly congruent with the effort required to form the original
energy - e.g., the formation of the first excited state energy and analytical
nuclear gradient of TD-DFT requires roughly double the cost of the formation of
only the first excited state energy, irrespective of the system size. In the
present work, we have nearly obtained this goal - in particular the invocation
of the Lagrangian formalism naturally led to a single set of response equations
to be solved, regardless of the number of nuclear or other derivative
perturbations applied, allowing e.g., for tractable production of nuclear
gradients with 264 perturbations in BChl-a dimer. However, close inspection of
the critical quantum portions of the gradient algorithm developed herein
indicates that the number of quantum observables required to form the gradient
does rise in a system-size-dependent manner over that required to form the
MC-VQE energy (assuming a gradient-based or Jacobi-1
style\cite{Parrish:2019:Jacobi} approach to converge the MC-VQE parameters). The
offending terms stem from the present manuscript's prescription to explicitly
evaluate the quantum Hessian quantities $\partial^2 \bar E / \partial \theta_{g}
\partial \theta_{g'}$ and $\partial^2 \bar E / \partial \theta_{g} \partial
\theta_{M}$ for the CP-SA-VQE LHS of Equation \ref{eq:vqe-hess} and the CP-CIS
RHS of Equation \ref{eq:cp-cis-rhs-2}, respectively. However, it is important to
recognize that these Hessian quantities are not intrinsically important
intermediates - all that we require is their action on certain trial vectors.
For instance, the mixed VQE-CIS angle Hessian term arises in a contribution to
the CP-CIS RHS where the salient intermediate is,
\begin{equation}
\label{eq:qprim-1}
\tau_{M}
\equiv
\sum_{g}
\tilde \theta_{g}^{\Theta}
\pdiff{^2 \bar E}{\theta_g \partial \theta_M}
\end{equation}
Additionally, if iterative linear equation solvers such as PCG,
DIIS,\cite{Pulay:1980:393,Pulay:1982:556} or GMRES are invoked (as is often done
in the solution of response equations in classical electronic structure
methods), the relevant quantity from the VQE-VQE angle Hessian term is,
\begin{equation}
\label{eq:qprim-2}
\tau_{g'}
\equiv
\sum_{g}
b_{g}
\pdiff{^2 \bar E}{\theta_g \partial \theta_{g'}}
\end{equation}
where $b_{g}$ is an arbitrary (classically known) trial vector. The contracted
quantum response primitives of Equations \ref{eq:qprim-1} and \ref{eq:qprim-2}
have thus far resisted an analytical linear-scaling tomography resolution, but
they resemble the gradients of directional derivatives of functions with
smoothness bounded by the overall sinusoidal support of the
$N_{\theta_g}$-dimensional VQE quantum circuit tomography. Therefore, it seems
likely that an accurate tomography approach could be used to resolve these
primitives with effort that is linear in $N_{\theta_g}$. Such an approach would
reduce the effort of forming the full analytical MC-VQE gradient to within a
constant prefactor of the underlying MC-VQE energy. Such methodology will be
considered in future work.

\section{Summary and Outlook}

We have explored the application of the Lagrangian formalism for analytical
derivative theory to the hybrid quantum/classical MC-VQE+AIEM method, with
specialization to the analytical nuclear gradient. In particular, we
demonstrated that this approach provides for a clean separation between the
quantum and classical portions of the algorithm, with a cost for the quantum
portion that is independent of the number of nuclei per AIEM monomer. Within the
quantum portions of the algorithms, the Lagrangian formalism initiated the
natural emergence of CP-SA-VQE and CP-CIS response equations which must be
solved to account for the definitional choices of the SA-VQE and CIS portions of
the MC-VQE wavefunction ansatz. We developed quadrature-based tomography
methodology to extract the needed gradient and Hessian matrix elements from
quantum circuit measurements, and have outlined a key primitive operation to
form contractions of the quantum Hessians with known trial vectors which could
potentially reduce the cost of the complete analytical MC-VQE gradient to a
constant scaling factor over the underlying MC-VQE energy. Within a trial BChl-a
dimer system with a truncated VQE entangler circuit, we validated the analytical
MC-VQE gradients by comparison to finite difference computations in the Pauli,
monomer-property, and nuclear perturbations, using an ideal quantum circuit
simulator. We then explored adiabatic excited state dynamics of the same system,
and demonstrated the importance of self-consistent nuclear gradients by showing
that Hellmann-Feynman-approximated MC-VQE gradients exhibit unphysical energy
excursions due to the missing response terms. In systems with more extensive VQE
entangler circuits, the absolute accuracy of MC-VQE will definitionally
increase, possibly leading to smaller response contributions that might be able to be
neglected in certain applications. However, it remains important to have fully
exact analytical derivative methodology to characterize the potential errors
arising from such situations. Moreover, if the proposed endeavor to develop
improved tomography methodology for the observation of quantum Hessian-vector
contractions succeeds, the overall analytical derivative methodology will cost
only a constant multiple of the underlying energy, so the use of full
response-including nuclear gradients should be vastly preferred to the use of
Hellmann-Feynman gradients.

The outlook for hybrid quantum/classical derivative methodology is, in our
opinion, highly promising. With the present manuscript, most of the known lore
on analytical derivative methodology that has been developed over the past few
decades has been translated from the classical to the quantum realm. There exist
a number of technically straightforward but highly compelling projects to extend
the present methodology to other derivative properties such as non-adiabatic
coupling vectors, polarizabilities, and optical response properties. There is
also a need to explore the use of this methodology within MC-VQE in Hamiltonians
beyond AIEM, e.g., in fermionic Hamiltonians. Finally, many practical issues
will have to be dealt with when making the jump from an ideal quantum circuit
simulator to noisy quantum hardware. These challenges are well worth tackling,
given the prospects of performing large-scale non-adiabatic dynamics simulations
with hybrid quantum/classical computing in the near future.

\textbf{Acknowledgements:} This material is based on work partially supported by
the U.S. Department of Energy, Office of Science, Office of Advanced Scientific
Computing Research (SciDAC)  program.

\textbf{Financial Disclosure:} TJM is a cofounder of \textsc{PetaChem LLC}.
RMP and PLM own stock/options in \textsc{QC Ware Corp.}

\textbf{Note Added:} As we were finalizing the numerical tests in the present
manuscript, two noteworthy manuscripts appeared on the arXiv that deal with the
topic of nuclear gradients of the energy in hybrid quantum/classical electronic
structure algorithms.\cite{Obrien:2019:grad,Mitarai:2019:grad} Both groups
demonstrate their respective methodologies in the context of nuclear derivatives
of the ground-state energy in the one-dimensional H$_2$ system - a system for
which the Hellmann-Feynman theorem holds for the first derivative for standard
VQE, and for which the one-dimensional nature of the system makes explicit
forward differentiation of wavefunction response contributions much more
tractable than in high-dimensional systems such as those encountered in
MC-VQE+AIEM.

% \bibliography{jrncodes.bib,refs.bib}
% \bibliographystyle{aip}

% ***** Figures and Tables *****

\newpage
\clearpage

\begin{table*}[h!]
\begin{center}
\caption{Comparison of Pauli-basis density matrices computed for BChl-a dimer
within the AIEM based on $\omega$PBE($\omega=0.3$)/6-31G*-D3. The upper portion
of the table shows the deviation between the analytical and finite-difference
formulations of the Pauli-basis density matrices, with a second-order symmetric
finite difference stencil with stepsize of $10^{-7}$ used for all quantities.
The bottom portion of the table shows the deviation between different methods
with analytical formulations of the Pauli-basis density matrices. In all cases,
the maximum absolute deviation across the ground-state and first-excited-state
Pauli density matrices is reported. The notation VQE(N,Y) indicates that VQE
response (the first argument) was not used while CIS response (the second
argument) was used.  For MC-VQE, two states were computed and the VQE entangler
circuit of Figure \ref{fig:dimer-schem} was used.} 
\label{tab:pauli_dm}
\begin{tabular}{l|lllll}
\hline \hline
              Method &     $\gamma_{\mathcal{X}_{A}}^{\Theta}$ &
$\gamma_{\mathcal{Z}_{A}}^{\Theta}$ &    $\Gamma_{\mathcal{XX}_{AA'}}^{\Theta}$ &
$\Gamma_{\mathcal{XZ/ZX}_{AA'}}^{\Theta}$  &    $\Gamma_{\mathcal{ZZ}_{AA'}}^{\Theta}$ \\
\hline \hline
\multicolumn{6}{c}{Analytical vs. Finite Difference} \\
\hline
            VQE(N,N) &                $1.2 \times 10^{-1}$ &                $1.2 \times 10^{-1}$ &                $5.7 \times 10^{-2}$ &                $6.1 \times 10^{-2}$  &                $4.0 \times 10^{-2}$ \\
            VQE(N,Y) &                $7.0 \times 10^{-2}$ &                $1.0 \times 10^{-2}$ &                $1.4 \times 10^{-2}$ &                $4.0 \times 10^{-2}$  &                $1.6 \times 10^{-2}$ \\
            VQE(Y,N) &                $5.0 \times 10^{-2}$ &                $1.3 \times 10^{-1}$ &                $7.1 \times 10^{-2}$ &                $5.0 \times 10^{-2}$  &                $2.4 \times 10^{-2}$ \\
            VQE(Y,Y) &               $3.0 \times 10^{-10}$ &               $2.4 \times 10^{-11}$ &               $2.6 \times 10^{-11}$ &               $2.1 \times 10^{-10}$  &               $5.0 \times 10^{-11}$ \\
                 FCI &               $1.3 \times 10^{-10}$ &               $1.7 \times 10^{-10}$ &               $1.2 \times 10^{-10}$ &               $9.1 \times 10^{-11}$  &               $7.6 \times 10^{-11}$ \\
                 CIS &               $1.5 \times 10^{-10}$ &               $1.5 \times 10^{-10}$ &               $3.1 \times 10^{-11}$ &               $1.5 \times 10^{-10}$  &               $9.6 \times 10^{-11}$ \\
\hline
\multicolumn{6}{c}{Method vs. Method (Analytical)} \\
\hline
             FCI-VQE &                $9.4 \times 10^{-2}$ &                $3.7 \times 10^{-2}$ &                $1.9 \times 10^{-1}$ &                $6.7 \times 10^{-2}$  &                $2.5 \times 10^{-2}$ \\
             FCI-CIS &                $2.3 \times 10^{-1}$ &                $1.5 \times 10^{-1}$ &                $1.6 \times 10^{-1}$ &                $2.4 \times 10^{-1}$  &                $3.7 \times 10^{-2}$ \\
             VQE-CIS &                $2.0 \times 10^{-1}$ &                $1.2 \times 10^{-1}$ &                $5.4 \times 10^{-2}$ &                $1.8 \times 10^{-1}$  &                $2.2 \times 10^{-2}$ \\
\hline \hline
\end{tabular}
\end{center}
\end{table*}

\begin{table*}[h!]
\begin{center}
\caption{Comparison of monomer property density matrices computed for BChl-a dimer
within the AIEM based on $\omega$PBE($\omega=0.3$)/6-31G*-D3. The upper portion
of the table shows the deviation between the analytical and finite-difference
formulations of the monomer property density matrices, with a second-order symmetric
finite difference stencil with stepsize of $10^{-7}$ used for the $\gamma$
quantities and with stepsize of $10^{-6}$ used for the $\vec \eta$ and $\vec
\zeta$ quantities. The bottom portion of the table shows the deviation between
different methods with analytical formulations of the monomer property density
matrices. In all cases, the maximum absolute deviation across the ground-state
and first-excited-state monomer property density matrices is reported. The
notation VQE(N,Y) indicates that VQE response (the first argument) was not used
while CIS response (the second argument) was used.  For MC-VQE, two states were
computed and the VQE entangler circuit of Figure \ref{fig:dimer-schem} was
used.} 
\label{tab:monomer_dm}
\begin{tabular}{l|lllllll}
\hline \hline
              Method &                  $\gamma_{H}^{A,\Theta}$ &                  $\gamma_{T}^{A,\Theta}$ &                  $\gamma_{P}^{A,\Theta}$ &               $\vec \eta_{H}^{A,\Theta}$ &               $\vec \eta_{T}^{A,\Theta}$ &               $\vec \eta_{P}^{A,\Theta}$ &              $\vec \zeta_{P}^{A,\Theta}$ \\
\hline
\multicolumn{8}{c}{Analytical vs. Finite Difference} \\
\hline
            VQE(N,N) &                     $6.0 \times 10^{-2}$ &                     $1.2 \times 10^{-1}$ &                     $6.0 \times 10^{-2}$ &                     $4.7 \times 10^{-4}$ &                     $5.0 \times 10^{-4}$ &                     $3.9 \times 10^{-4}$ &                     $3.1 \times 10^{-4}$ \\
            VQE(N,Y) &                     $5.2 \times 10^{-3}$ &                     $7.0 \times 10^{-2}$ &                     $5.2 \times 10^{-3}$ &                     $7.3 \times 10^{-5}$ &                     $1.9 \times 10^{-4}$ &                     $3.7 \times 10^{-4}$ &                     $1.3 \times 10^{-4}$ \\
            VQE(Y,N) &                     $6.3 \times 10^{-2}$ &                     $5.0 \times 10^{-2}$ &                     $6.3 \times 10^{-2}$ &                     $4.4 \times 10^{-4}$ &                     $4.5 \times 10^{-4}$ &                     $3.8 \times 10^{-4}$ &                     $3.0 \times 10^{-4}$ \\
            VQE(Y,Y) &                    $2.0 \times 10^{-10}$ &                    $1.6 \times 10^{-10}$ &                    $1.7 \times 10^{-10}$ &                    $4.0 \times 10^{-11}$ &                    $3.3 \times 10^{-11}$ &                    $3.9 \times 10^{-11}$ &                    $2.1 \times 10^{-11}$ \\
                 FCI &                    $8.4 \times 10^{-11}$ &                    $1.3 \times 10^{-10}$ &                    $7.3 \times 10^{-11}$ &                    $1.3 \times 10^{-11}$ &                    $1.4 \times 10^{-11}$ &                    $2.3 \times 10^{-11}$ &                    $3.4 \times 10^{-11}$ \\
                 CIS &                    $4.0 \times 10^{-11}$ &                    $1.5 \times 10^{-10}$ &                    $6.4 \times 10^{-11}$ &                    $2.6 \times 10^{-11}$ &                    $9.0 \times 10^{-12}$ &                    $1.6 \times 10^{-11}$ &                    $1.6 \times 10^{-11}$ \\
\hline
\multicolumn{8}{c}{Method vs. Method (Analytical)} \\
\hline
             FCI-VQE &                     $1.9 \times 10^{-2}$ &                     $9.4 \times 10^{-2}$ &                     $1.9 \times 10^{-2}$ &                     $2.3 \times 10^{-4}$ &                     $1.2 \times 10^{-3}$ &                     $4.6 \times 10^{-4}$ &                     $1.2 \times 10^{-3}$ \\
             FCI-CIS &                     $7.7 \times 10^{-2}$ &                     $2.3 \times 10^{-1}$ &                     $7.7 \times 10^{-2}$ &                     $4.0 \times 10^{-4}$ &                     $1.2 \times 10^{-3}$ &                     $1.6 \times 10^{-3}$ &                     $1.7 \times 10^{-3}$ \\
             VQE-CIS &                     $5.8 \times 10^{-2}$ &                     $2.0 \times 10^{-1}$ &                     $5.8 \times 10^{-2}$ &                     $3.3 \times 10^{-4}$ &                     $8.8 \times 10^{-4}$ &                     $1.3 \times 10^{-3}$ &                     $1.1 \times 10^{-3}$ \\
\hline \hline
\end{tabular}
\end{center}
\end{table*}

\begin{table*}[h!]
\begin{center}
\caption{Comparison of nuclear gradients computed for BChl-a dimer
within the AIEM based on $\omega$PBE($\omega=0.3$)/6-31G*-D3. The upper portion
of the table shows the deviation between the analytical and finite-difference
formulations of the monomer property density matrices, with a second-order symmetric
finite difference stencil with stepsize of $0.002$ \AA{}. The bottom portion of
the table shows the deviation between different methods with analytical
formulations of the nuclear gradients. In all cases, the maximum
absolute deviation across the ground-state and first-excited-state monomer
nuclear gradients is reported. The notation VQE(N,Y) indicates that VQE
response (the first argument) was not used while CIS response (the second
argument) was used.  For MC-VQE, two states were computed and the VQE entangler
circuit of Figure \ref{fig:dimer-schem} was used. The notation Mg$_{A}^{20}$
indicates that the comparison is over the $x$, $y$, and $z$ components of the
nuclear gradient on the Mg atom at index 20 (zero-based) on monomer $A$.}  
\label{tab:nuclear_dm}
\begin{tabular}{l|lllll}
\hline \hline
              Method &                            Mg$_{A}^{20}$ &                             O$_{A}^{41}$ &                             N$_{A}^{16}$ &                             C$_{B}^{22}$ &                             H$_{B}^{31}$ \\
\hline
\multicolumn{6}{c}{Analytical vs. Finite Difference} \\
\hline
            VQE(N,N) &                     $7.5 \times 10^{-4}$ &                     $1.4 \times 10^{-3}$ &                     $1.1 \times 10^{-3}$ &                     $1.8 \times 10^{-3}$ &                     $2.7 \times 10^{-5}$ \\
            VQE(N,Y) &                     $4.0 \times 10^{-4}$ &                     $7.1 \times 10^{-4}$ &                     $1.3 \times 10^{-3}$ &                     $1.1 \times 10^{-3}$ &                     $2.0 \times 10^{-5}$ \\
            VQE(Y,N) &                     $3.5 \times 10^{-4}$ &                     $7.1 \times 10^{-4}$ &                     $1.1 \times 10^{-3}$ &                     $2.9 \times 10^{-3}$ &                     $4.6 \times 10^{-5}$ \\
            VQE(Y,Y) &                     $6.3 \times 10^{-7}$ &                     $5.5 \times 10^{-6}$ &                     $2.2 \times 10^{-6}$ &                     $2.5 \times 10^{-6}$ &                     $1.3 \times 10^{-6}$ \\
                 FCI &                     $6.3 \times 10^{-7}$ &                     $5.5 \times 10^{-6}$ &                     $2.4 \times 10^{-6}$ &                     $2.5 \times 10^{-6}$ &                     $1.3 \times 10^{-6}$ \\
                 CIS &                     $6.2 \times 10^{-7}$ &                     $5.5 \times 10^{-6}$ &                     $1.9 \times 10^{-6}$ &                     $2.5 \times 10^{-6}$ &                     $1.3 \times 10^{-6}$ \\
\hline
\multicolumn{6}{c}{Method vs. Method (Analytical)} \\
\hline
             FCI-VQE &                     $5.7 \times 10^{-4}$ &                     $9.6 \times 10^{-4}$ &                     $1.0 \times 10^{-3}$ &                     $5.8 \times 10^{-4}$ &                     $2.2 \times 10^{-5}$ \\
             FCI-CIS &                     $1.7 \times 10^{-3}$ &                     $2.9 \times 10^{-3}$ &                     $5.3 \times 10^{-3}$ &                     $7.6 \times 10^{-4}$ &                     $2.0 \times 10^{-5}$ \\
             VQE-CIS &                     $1.5 \times 10^{-3}$ &                     $2.5 \times 10^{-3}$ &                     $4.3 \times 10^{-3}$ &                     $3.2 \times 10^{-4}$ &                     $1.3 \times 10^{-5}$ \\
\hline \hline
\end{tabular}
\end{center}
\end{table*}

\begin{table*}[h!]
\begin{center}
\caption{
Comparison of nuclear gradients computed for BChl-a hexamer ($N=6$) within AIEM
based on $\omega$PBE($\omega=0.3$)/6-31G*-D3. The upper portion of the table
shows the errors of the MC-VQE energy and oscillator strength observables vs.
FCI. The lower portion of the table shows the maximum absolute deviations
between various analytical gradient methods for the first excited state nuclear
energy gradient. The first column of data shows the results from a truncated VQE
entangler with a single $R_y$ gate at the end of each qubit wire, analogous to
the truncated VQE entangler used for the BChl-a dimer example in the rest of the
manuscript. The second column of data shows the results from a full $SO(4)$
entangler of Equation \ref{eq:vqe_entangler_merged}.
}
\label{tab:large}
\begin{tabular}{l|ll}
\hline \hline
Method & 1-Layer & 5-Layer \\
\hline
\multicolumn{3}{c}{Accuracy Characteristics} \\
\hline
$E^{0}$                     & $5.4 \times 10^{-3}$ & $7.6 \times 10^{-4}$ \\
$E^{1}$                     & $3.9 \times 10^{-3}$ & $5.7 \times 10^{-4}$ \\
$\Delta E^{0\rightarrow 1}$ & $1.4 \times 10^{-3}$ & $1.9 \times 10^{-4}$ \\
$O^{0 \rightarrow 1}$       & $1.5 \times 10^{-4}$ & $5.9 \times 10^{-5}$ \\
\hline
\multicolumn{3}{c}{Gradient Deviations} \\
\hline
VQE(Y,Y)-VQE(N,N)           & $1.2 \times 10^{-3}$ & $1.7 \times 10^{-3}$ \\
VQE(Y,Y)-VQE(N,Y)           & $2.4 \times 10^{-3}$ & $1.5 \times 10^{-3}$ \\
VQE(Y,Y)-VQE(Y,N)           & $1.2 \times 10^{-3}$ & $7.5 \times 10^{-4}$ \\
FCI-VQE(Y,Y)                & $4.9 \times 10^{-3}$ & $1.1 \times 10^{-3}$ \\ 
FCI-CIS                     & $8.5 \times 10^{-3}$ & $8.5 \times 10^{-3}$ \\ 
VQE(Y,Y)-CIS                & $3.6 \times 10^{-3}$ & $7.5 \times 10^{-3}$ \\ 
\hline \hline
\end{tabular}
\end{center}
\end{table*}

\end{document}